\documentclass[11pt,a4paper]{article}
\pdfoutput=1

\usepackage{ifpdf}
\usepackage{jheppub}
\usepackage{rotating}
\usepackage[bb=boondox]{mathalfa}
\usepackage{comment}
\usepackage{tikz-cd}
\usepackage{slashed}
\usepackage{graphicx} 
\usepackage{mathrsfs}
\usepackage{amsmath,amsfonts,amssymb}
\usepackage{mathtools}
\usepackage{dcolumn}
\usepackage{bm}
\usepackage{hyperref}
\usepackage[mathlines]{lineno}
\usepackage{xcolor}
\usepackage{float}
\usepackage{blindtext}
\usepackage{titlesec}
\title{Sections and Chapters}
\usepackage[toc,page]{appendix}
\usepackage{xcolor}
\usepackage{graphicx}
\graphicspath{ {./images/} }
\usepackage[thinc]{esdiff}
\usepackage[euler]{textgreek}
\usepackage{braket}
\usepackage{physics}
\usepackage{xfrac}
\usepackage{soul}
\usepackage{stmaryrd}
\usepackage{pifont}
\usepackage{orcidlink}
\usepackage{amssymb}
\usepackage{multirow}

\usepackage[normalem]{ulem}
\usepackage{color}
\definecolor{darkgreen}{cmyk}{1,0,1,0.4}
\definecolor{darkcyan}{cmyk}{1,0,0,0.4}

\def\barr{\begin{array}}
\def\earr{\end{array}}
\def\dis{\displaystyle}
\definecolor{deepbrown}{RGB}{210, 180, 140}

\makeatletter
\renewcommand{\fnum@table}{\textbf{\tablename~\thetable}}
\renewcommand{\fnum@figure}{\textbf{\figurename~\thefigure}}
\makeatother

\allowdisplaybreaks
\title
{ Solving Cosmological Puzzles using Finite Temperature $\nu$SMEFT} 

\author[a]{Debajyoti Choudhury,\,\orcidlink{0000-0002-8124-0043}\,}  
\author[b]{\!\! Jaydeb Das,\,\orcidlink{0000-0001-6335-9377}\,}
\author[a,d]{Tripurari Srivastava\,\orcidlink{0000-0001-6856-9517}\,}  
\affiliation[a]{Department of Physics and Astrophysics, University of Delhi, Delhi-110007, Delhi, India,} 
\affiliation[b]{Department of Physics, Indian Institute of Technology Guwahati,
North Guwahati, Assam-781039, India,}  
\affiliation[d]{Institute of Particle Physics and Key Laboratory of Quark and Lepton Physics (MOE),
Central China Normal University, Wuhan, Hubei 430079, China}

\emailAdd{debchou.physics@gmail.com}
\emailAdd{jaydebphys@rnd.iitg.ac.in}
\emailAdd{tripurarisri022@gmail.com}
\begin{document}
\abstract
{ We study a minimal framework that naturally yields viable Dark
Matter, a strong first-order electroweak phase transition and
low-scale resonant leptogenesis.  Augmenting the Standard Model with
three heavy Majorana neutrinos, we study the corresponding
neutrino-extended Standard Model Effective Field Theory, including
operators upto mass-dimension six.  The pure Higgs operator provides
the dominant enhancement required for a strong first-order electroweak
phase transition, while the remaining operators yield subleading
effects consistent with electroweak precision constraints. The signal
for the stochastic gravitational-wave background is dominated by sound
waves in the plasma, with magnetohydrodynamic turbulence providing a
subleading contribution.  Low-scale resonant leptogenesis is realized
through tiny mass splittings among quasi-degenerate heavy neutrinos,
dynamically generated in the symmetric phase by the combined effect of
one-loop RG-induced corrections and finite-temperature
contributions. Solving the Boltzmann equations, we show that the
observed baryon asymmetry of the Universe can be reproduced while
remaining consistent with neutrino oscillation data and
charged-lepton-flavor-violation constraints. One of the
heavy neutrinos is stabilized by a discrete symmetry thereby acting as
a fermionic dark matter candidate. Its interactions with the Standard
Model arise from dimension-five and dimension-six effective operators,
leading to viable annihilation, elastic scattering, and indirect
detection phenomenology compatible with current experimental
bounds. The dark matter sector remains decoupled from the dynamics of
the electroweak phase transition and leptogenesis, allowing all three
phenomena to be consistently realized within a unified effective field
theory framework.}

\maketitle
\section{Introduction}
The observed imbalance between matter and antimatter in the Universe
constitutes one of the most compelling indications that the Standard
Model (SM) of particle physics is incomplete. The baryon asymmetry is
conventionally quantified by the ratio of the baryon number density to
the entropy density, which has been precisely determined from Big Bang
Nucleosynthesis and measurements of the cosmic microwave background by
the Planck mission~\cite{Planck:2018vyg},
\[
Y_B \equiv \frac{n_B}{s} = (8.65 \pm 0.09)\times10^{-11}.
\]
Any successful dynamical explanation of this asymmetry must
simultaneously satisfy the Sakharov
conditions~\cite{AndreiDSakharov_1991}: violation of baryon number,
violation of charge conjugation and charge–parity symmetries, and a
departure from thermal equilibrium.

In addition to the baryon asymmetry, the existence of non-baryonic
dark matter provides further evidence for physics beyond the Standard
Model~\cite{Bertone:2004pz}.  
Cosmological and astrophysical observations indicate that dark
matter constitutes about $26\%$ of the total energy density of the
Universe~\cite{Planck:2018vyg}, yet the SM contains no viable dark matter
candidate. 
It is, thus, an
  intriguing possibility that the mechanism responsible for
baryogenesis is linked to a more general
extension of the SM that also addresses the dark matter problem.

Although the SM satisfies the Sakharov conditions
in principle, it fails quantitatively. Baryon number violation arises
through nonperturbative sphaleron processes (which are extremely
suppressed unless the temperature is above the electroweak scale), and
charge–parity violation is present in the Cabibbo–Kobayashi–Maskawa
(CKM) matrix, but the magnitude of this violation is far too small to
account for the observed asymmetry. More importantly, nonperturbative
lattice studies have established that, given the physical Higgs boson
mass, the electroweak phase transition in the SM can only be a smooth
crossover rather than a genuine first-order
transition~\cite{Kajantie:1996mn,Csikor:1998eu,DOnofrio:2015gop}. As a
result, the departure from thermal equilibrium required for
electroweak baryogenesis is absent too.

A wide class of extensions of the SM can modify the finite-temperature
Higgs potential and render the electroweak phase transition (EWPT)
strongly first order. This possibility has been extensively explored
through the introduction of additional scalar degrees of freedom,
including singlet extensions with discrete symmetries such as $Z_2$
and
$Z_3$~\cite{Barger:2007im,Espinosa:2011ax,Noble:2007kk,Cline:2012hg,Alanne:2014bra,Cline:2013gha,Chiang:2019oms,Kannike:2019mzk,Srivastava:2025oer,Chaudhuri:2022sis,Das:2026zuo,Chakrabarty:2022yzp}. More
recently, it has been shown that higher-dimensional operators (HDOs)
within the Standard Model Effective Field Theory (SMEFT) framework can
also strengthen the electroweak phase transition without introducing
light new
particles~\cite{Camargo-Molina:2021zgz,Hashino:2022ghd,Oikonomou:2024jms}.

The absence of direct evidence for new particles at the Large Hadron
Collider (LHC) motivates the possibility that additional degrees of
freedom reside at energy scales above the electroweak scale. In this
context, effective field theory (EFT) provides a systematic and
model-independent framework to capture the low-energy effects of heavy
new physics through higher-dimensional operators constructed from SM
fields. In particular, dimension-six operators involving the Higgs
field can significantly alter the thermal Higgs potential while
remaining consistent with precision electroweak measurements, thereby
enabling a strong first-order electroweak phase transition
(SFOEWPT)~\cite{Anderson:1991zb,Zhang:1992fs,Camargo-Molina:2021zgz,Hashino:2022ghd,Oikonomou:2024jms}.

An alternative and well-motivated mechanism for generating the baryon
asymmetry is
leptogenesis~\cite{Fukugita:1986hr,Pilaftsis:1997jf,Buchmuller:2005eh,Davidson:2008bu,Chakraborty:2022pcc,Mohanty:2019drv,Bhattacharya:2023kws,Bhandari:2025ufp}. In
this scenario, a lepton asymmetry is produced through
lepton-number-violating interactions involving heavy Majorana
neutrinos and is subsequently converted into a baryon asymmetry by the
electroweak sphaleron
processes~\cite{Rubakov:1996vz,Klinkhamer:1984di,Manton:1983nd}. Leptogenesis
naturally links the origin of the baryon asymmetry with the generation
of neutrino masses via the seesaw mechanism.

Within the SMEFT framework, lepton number violation first appears at
dimension five through the Weinberg operator. The constraints on the
neutrino masses ($\Delta m^2$ from the oscillation experiments~\cite{Esteban:2020cvm} and
$\sum_i m_{\nu_i}$ from cosmology~\cite{Planck:2018vyg}) together suggest that the effective
mass scale of the operator is as high as $\sim 10^{14}$~GeV. Lowering
this scale to ${\cal O}({\rm TeV})$---thereby bringing the new physics
into the observable arena---would need to ensure that this operator first appears only at a sufficiently high loop order.

Low-scale leptogenesis relies, instead, on a resonant enhancement of
the charge–parity asymmetry, which occurs when two or more heavy
neutrinos are nearly degenerate in mass. Tiny mass splittings between
these states lead to a resonant enhancement of the asymmetry in their
decays, allowing for successful leptogenesis at the TeV scale with
naturally small Yukawa couplings. This mechanism, known as resonant
leptogenesis, provides a theoretically consistent and
phenomenologically viable route to baryogenesis at accessible energy
scales. While the required smallness of
  the splitting might seem artificial, we demonstrate that it can be
  engendered by tiny finite-temperature corrections to the
  self-energies of heavy neutrinos that are degenerate owing to
  symmetries.

In this work, we extend the SM by introducing three generations of
right-handed Majorana neutrinos and include operators up to dimension
six, forming a neutrino-extended SMEFT. The strong first-order
electroweak phase transition is primarily driven by dimension-six
Higgs operators.  In the quasi-degenerate regime considered here, the
tiny mass splitting required for resonant leptogenesis is
already realized even in the symmetric phase
through the combined effect of one-loop renormalization-group running
and finite-temperature Yukawa corrections. After electroweak symmetry
breaking, diagonalization of the full neutrino mass matrix gives an
additional vacuum contribution to the heavy-neutrino spectrum. The
contribution from the dimension-5 operator NNHH remains
subleading.  In our setup, only two of the singlet fermions actively
participate in the seesaw mechanism and resonant leptogenesis. The
third singlet state is isolated by a discrete symmetry, which forbids
its mixing with the active neutrinos and renders it stable on
cosmological timescales. A strong first-order electroweak phase
transition also gives rise to a stochastic background of gravitational
waves through bubble nucleation and plasma
dynamics~\cite{Kamionkowski:1993fg,Ellis:2018mja,Croon:2018erz,Beniwal:2018hyi,Huang:2017laj,Hashino:2018zsi,Demidov:2017lzf,Mazumdar:2018dfl,Kobakhidze:2016mch,Kobakhidze:2017mru,Dev:2016feu}. The
breakthrough detection of gravitational waves by
LIGO~\cite{LIGOScientific:2016aoc} has opened a new observational
window into early-Universe dynamics, motivating the study of
electroweak-scale gravitational waves as a complementary probe of
physics beyond the SM. The resulting signal is potentially observable
by future space-based and ground-based gravitational-wave
interferometers such as LISA~\cite{eLISA:2013xep},
DECIGO~\cite{Kawamura:2011zz}, BBO~\cite{Corbin:2005ny},
TAIJI~\cite{Gong:2014mca}, and TianQin~\cite{TianQin:2015yph},
providing an independent probe of the mechanism responsible for
baryogenesis. Within the same effective field theory framework, the
stabilized third singlet fermion provides a natural fermionic dark
matter
candidate~\cite{Blennow:2019fhy,Bandyopadhyay:2022tsf,Criado:2021trs,Matsumoto:2014rxa}.
And while it does not have a leading effect on electroweak phase
transition or leptogenesis, its interactions with the Standard Model
courtesy higher-dimensional operators, leading to testable dark matter
annihilation and scattering signatures.

This paper is organized as follows. In Section~\ref{sec:model}, we
introduce the theoretical framework and effective Lagrangian. The
one-loop effective potential at finite temperature is computed in
Section~\ref{sec:ftep}, and the electroweak phase transition is
analyzed in Section~\ref{sec:FOPT}. The associated gravitational-wave
signatures are discussed in Section~\ref{sec:gw}. Neutrino mass
generation and parametrization are presented in
Section~\ref{sec:neumass}, followed by constraints from charged-lepton
flavor violation in Section~\ref{sec:clfv}. Leptogenesis is studied in
Section~\ref{sec:leptogenesis}, with numerical results shown in
Section~\ref{sec:numerical}. Dark matter phenomenology is discussed in Section~\ref{sec:dm}.
We summarize and conclude in
Section~\ref{sec:concl}.

\section{Theoretical framework} \label{sec:model}
With one of our main goals being to induce the observed
matter-antimatter asymmetry through leptogenesis, we must enhance the
leptonic sector of the SM. Aiming for a minimal augmentation, the only
change in the field content is the addition of three SM-singlet
right-handed neutrinos $N_i$. As to any other new physics above the
$N_i$ mass scale, we remain agnostic, parametrizing our ignorance by
the inclusion of higher-dimension operators. While the latter can be
of diverse character, we largely concentrate on the sector that
can have the most telling effect on a possible phase transition,
namely couplings involving the SM Higgs field $H$. Restricting
ourselves to terms upto dimension-6, we have then
 \begin{eqnarray}
   \mathcal{L}&&\supset (D_\mu H )(D^\mu H)  -\mu_H^2 (H^\dagger H)
   - \lambda_H (H^\dagger H)^2
   +  \mathcal{L}_{HN}
   +\frac{1}{\Lambda^2}\sum_i\mathcal{C}_i \mathcal{O}_i  \, ,
   \label{eq:lagr}
 \end{eqnarray}
 with
 \begin{eqnarray}
   \mathcal{L}_{HN} = \overline{N_i} (i\slashed{\partial}) N_i
   - \left(\frac{m_{R_{ij}}}{2}\overline{N_i^c} N_j
   +\mathcal{Y}_{D_{ij}} \overline{L_i} \tilde{H} N_j
  +  \frac{c_{{NH}_{ij}}}{2\Lambda} (\overline{N_i^c} N_j) (H^\dagger H) +\text{h.c} \right),
   \label{eq:lag_H_N}
  \end{eqnarray}
 where $L_i$ are the SM lepton doublets, and we have exploited the fact that the $N_i$ are right-handed Majorana fermions. While the
 Yukawa coupling matrix $\mathcal{Y}_{D}$ is, in general, a complex one, the mass matrix $m_R$, of course, needs to be symmetric, and,
 without any loss of generality, can be taken to be diagonal. In addition, we impose a discrete $\mathbb{Z}_2$ symmetry under which
 $N_3$, is odd, while all other fields in the theory are even. This symmetry forbids not only the Yukawa interaction $\overline{L}\tilde
 H N_3$, but also assures the stability of $N_3$, thereby allowing it to play the role of the dark matter. We defer a detailed discussion of the dark matter sector and its phenomenology to a later section.

 We will argue later for the two $Z_2$-even heavy neutrinos being very close in mass or even degenerate\footnote{While this could, in principle, be motivated by imposing some symmetries, we choose not to do so. Rather, we posit it as a way of reducing the number of free parameters in the theory, as also to facilitate efficient leptogenesis.} with $N_3$ being somewhat heavier. We, however, make no corresponding demand on the Wilson coefficient matrix $c_{NH}$, with a consequent small lifting of the mass degeneracy.
 
While the only possible dimension-5 terms have been explicitly listed
in Eq.(\ref{eq:lag_H_N}), the $\mathcal{O}_i$ in Eq.(\ref{eq:lagr})
are the dimension six SMEFT operators with $\mathcal{C}_i$ being the
corresponding Wilson coefficients (WCs).  The pure bosonic operators
relevant to the Higgs effective action
are~\cite{Hashino:2022ghd,Grzadkowski:2010es}.
\begin{eqnarray}
\mathcal{O}_{H}= \left(H^\dagger H\right)^3, \quad \mathcal{O}_{HD}= \left(H^\dagger \mathcal{D}_\mu H\right)^\ast \left(H^\dagger \mathcal{D}^\mu H\right),\quad \mathcal{O}_{H\Box}= (H^\dagger H)\Box (H^\dagger H).
\end{eqnarray}
whereas those involving both the Higgs fields and up-type
quarks are
\begin{equation}
  \barr{rclcrcl}
  \dis (\mathcal{O}_{uH})_{ij}&=& \dis
  (H^{\dagger}H)(\bar{q}_i u_j \tilde{H}) &\qquad \quad&
   \dis (\mathcal{O}_{Hq}^{(1)})_{ij}&=& \dis(H^{\dagger}i\overleftrightarrow{D}_{\mu}H)(\bar{q}_i \gamma^{\mu}q_j),
    \\[2ex]
 \dis (\mathcal{O}_{Hq}^{(3)})_{ij}&=& \dis(H^{\dagger}i\overleftrightarrow{D}_{\mu}^IH)(\bar{q}_i \tau^I \gamma^{\mu}q_j),& &
   \dis (\mathcal{O}_{Hu})_{ij}&=& \dis(H^{\dagger}i\overleftrightarrow{D}_{\mu}H)(\bar{u}_i  \gamma^{\mu}u_j),
   \earr
   \label{op_boson_top}
\end{equation}
with the two-sided derivative being defined by
\begin{align}
    H^{\dagger}\overleftrightarrow{D}^I_{\mu} H=H^{\dagger} \tau^I D_{\mu} H-(D_{\mu}H)^{\dagger}\tau^I H.
\end{align}
Here, $q_i$ are the $SU(2)_L$ quark doublets, $u_j$ the right-handed
up-type quarks, and $\tau^I$ the Pauli matrices. Of the four sets of
operators listed in Eq.(\ref{op_boson_top}), the last three would turn
out to have vanishing contributions (at the leading order) to the
effective potential for the neutral scalar and, hence, are
largely irrelevant to our analysis. Concentrating on the most
relevant piece, namely the top-sector, the $(\mathcal{O}_{uH})_{ij}$
operator gives a field dependent mass to the top quark, {\em viz.},
\begin{align}
        m_t(h,C_{tH})\equiv \frac{y_t h}{\sqrt{2}} - \frac{C_{tH} h^3}{2\sqrt{2}\Lambda^2}
\end{align}
where $y_t$ is the tree-level top Yukawa coupling within
  the SM, and $h$ is the value of the neutral scalar
field\footnote{Since we would be interested in finite-temperature
effects, $h$ is not necessarily the same as $v \approx 246$~GeV.}. Since only the top quark contribution is relevant for the analysis of EWPT, we henceforth denote the Wilson coefficient $C_{uH}$ by $C_{tH}$.

It is easy to see that the tree-level contributions
of the operators $\mathcal{O}_{HD}$ and
$\mathcal{O}_{H\Box}$ to the Higgs Lagrangian are
\begin{eqnarray}
  \mathcal{O}_{HD} \supset \frac{1}{4} \frac{\mathcal{C}_{HD}}{\Lambda^2} h^2 \left(\partial_\mu h\right)^2\qquad {\rm and} \qquad
  \mathcal{O}_{H\Box} \supset-\frac{\mathcal{C}_{H\Box}}{\Lambda^2} h^2 \left(\partial_\mu h\right)^2 \ .
\end{eqnarray}
Thus, the leading contributions from the dimension-6 SMEFT
operators to the Higgs effective action can be summarised as
\begin{eqnarray}
    \delta \mathcal{L}_{\rm SMEFT}= \frac{\mathcal{C}_{\rm kin}}{\Lambda^2}  h^2 \left(\partial_\mu h\right)^2 + \frac{1}{8}\frac{\mathcal{C}_H}{\Lambda^2} h^6,\qquad {\rm where} \qquad \mathcal{C}_{\rm kin} = \frac{1}{4} \mathcal{C}_{HD} - \mathcal{C}_{H\Box}\ .
    \label{eq:ckin}
\end{eqnarray}

In addition, the Wilson coefficients \(C_{HD}\), \(C_{H\Box}\),
\(C_{tH}\), and \(C_H\) do modify Higgs couplings to the gauge bosons as well as to fermions, thereby
affecting processes such as \(t\bar t H\) production, Higgs--gauge
boson interactions, and electroweak precision observables. The
constraints from such observables~\cite{Ellis:2020unq} are explicitly
taken into account in our numerical analysis.
   
It is useful, at this stage, to effect a field redefinition to render
the kinetic term canonical, namely
\begin{eqnarray}
    h \rightarrow h - \frac{1}{3} \frac{\mathcal{C}_{\rm kin}}{\Lambda^2} h^3 \ .
\end{eqnarray}
Although the field shift induces a correction in
\(\mathcal{L}_{HN}\), this effect is suppressed by an additional
factor of \(v^2/\Lambda^2\) and thus remains well within the allowed fit
ranges. 
 In terms of the new field, we have
\begin{eqnarray}
    \mathcal{L}_{\rm Higgs} = \frac{1}{2}\left(\partial_\mu h\right)^2 - V_0(h) + \mathcal{O}\left(\frac{1}{\Lambda^{4}}\right),
\end{eqnarray}
where the tree-level Higgs potential is now given by
\begin{eqnarray}\label{eq:pottree}
    V_0 (h) = \frac{a_2}{2} h^2 + \frac{a_4}{4} h^4 + \frac{a_6}{6}  h^6, 
\end{eqnarray}
with 
\begin{eqnarray}\label{eq:input}
    a_2 &=& \mu_H^2, \quad a_4 = \lambda_H -\frac{4}{3}\frac{\mathcal{C}_{\rm kin}}{\Lambda^2}\mu_H^2, \quad a_6 = -\frac{3}{4}\frac{\mathcal{C}_H}{\Lambda^2} - 2 \lambda_H \frac{{\cal C}_{\rm kin}}{\Lambda^2} \ ,
\end{eqnarray}
and stability of the tree-level potential demands that $a_6 >
0$ or, in other words, ${\cal C}_H + 8 {\cal C}_{\rm kin} <
0$. Experimentally, $a_6$ could be determined by measuring the Higgs
self couplings, {\em e.g.}  the triple-Higgs coupling at the
LHC. Pending this, the parameters could be related to the measurables
$v$ and $m_h$ through the potential minimization conditions, yielding,
at the tree level,
\begin{eqnarray}\label{eq:const}
    a_2 &=& \frac{1}{2} (-m_h^2 + 2 a_6 v^4 ), \quad a_4 = -\frac{4 a_6 v^4-m_h^2}{2 v^2}\ ,
\end{eqnarray}
and leaving one of $a_{2,4,6}$ as undetermined.
  For our numerical analysis, the input parameters are, then, the Wilson coefficients $C_H,\, C_{\rm
    kin},\, C_{tH}$, apart from the mass of Higgs $m_h$ and its 
  VEV $v$ at
$T=0$.
\section{The effective potential at nonzero temperatures}
\label{sec:ftep}
The effective potential, understandably,
depends on multiple parameters in the theory. It is convenient to
describe the latter in terms of measurables such as physical
masses. With the inclusion of the higher-dimensional terms in the
Lagrangian (especially, the operators $\mathcal{O}_{HD}$ and
$\mathcal{O}_{H\Box}$), these relations change from those within the
SM, and we have, instead, for the bosonic fields
\begin{equation}\label{eq:mass0boson}
  \barr{rclcrcl}
  &&m_W^2(h)  =  \dis \frac{1}{4} g^2 h^2
  \left(1-2\frac{\mathcal{C}_{\rm kin} h^2}{3\Lambda^2}\right),
  & \qquad &
&& m_Z^2(h) =  \dis
  \frac{g^2 + g^{' 2}}{4}h^2
  \left(1-2\frac{\mathcal{C}_{\rm kin} h^2}{3\Lambda^2}\right), \\[2ex]
&&m_h^2 (h)  =   \dis a_2 + 3  a_4 h^2 + 5 a_6 h^4 , & & 
&&m_{\chi_i}^2(h)  =  \dis a_2 +  a_4 h^2 +  a_6 h^4 ,
\earr
\end{equation}
where we have denoted the expectation value of the neutral scalar
field by $h$ (since we are still talking of $T = 0$, this is to be
identified with $v \approx 246$ GeV). As for the fermion fields, only
the top-quark and the right-handed neutrinos are of relevance.  While
the tree-level values are easy to read off from the effective
Lagrangian, note that these do receive corrections from the operators
$\mathcal{O}_{HD}$ and $\mathcal{O}_{H\Box}$. To ${\cal
  O}(\Lambda^{-2})$, these masses now read\footnote{ In principle, the
field dependent heavy neutrino masses follow from diagonalizing the
full $h$ dependent mass matrix $m_{N_{ij}}(h)= m_{R_{ij}}+ c_{NH_{ij}}
h^2 /(2\Lambda)$.  In our setup, the off diagonal entries of $c_{NH}$
are purely imaginary and numerically small, so their effect on the
field dependent eigenvalues is negligible for the electroweak phase
transition. We therefore approximate $M_{i}(h)\simeq
m_{R_i}+\mathrm{Re}(c_{NH_{ii}})h^2 /(2\Lambda)$ when evaluating the
thermal potential.}
\begin{equation}\label{eq:mass0}
  \barr{rcl}
&&m_t(h) =  \dis \frac{y_t h}{\sqrt{2}}\left(1-\frac{\mathcal{C}_{\rm kin} h^2}{3\Lambda^2}\right)-\frac{C_{tH} h^3}{\sqrt{8} \Lambda^2}, \\[2ex]
&&M_{i}(h)  \simeq \dis m_{R_i} + \frac{\text{Re}(c_{{NH}_{ii}})}{2\Lambda}\, h^2  \ .
  \earr
\end{equation}
It is obvious that placing $h = v$ (as is the case at $T=0$) would
allow us to invert the relations and determine the Wilson coefficients
in terms of the measurables.

Given the above, we may now compute the effective potential at nonzero
temperatures. For our purposes, it suffices to consider only the
one-loop corrections and it is instructive to start with the
(zero-temperature) Coleman-Weinberg (CW) potential. Employing the
$\overline{MS}$ regularization scheme, this can be written
as~\cite{Coleman:1973jx,Wainwright:2011kj,Blinov:2015vma,Aoki:2021oez}:
\begin{equation}\label{eq:CW}
  V^{\text{CW}}_{\rm 1-loop}(h)=\frac{1}{64\pi^2}
  \sum_i (\pm g_i) m_i^4(h)\left[
    \ln \left(\frac{m_i^2(h)}{\mu^2}\right)-f_i\right],
\end{equation}
where $\mu$ is the renormalization scale of the theory, and the $\pm$  refers to whether the particle in the loop is a boson(fermion).
The degrees of
freedom $g_i$ equate $2 N_c$ for a Weyl fermion ($N_c$ being the number
of colours), $4 N_c$ for a Dirac fermion and $1$ for a scalar or for
each polarization component of a gauge boson\footnote{Clearly, the
$W^\pm$ consitute individual entities while, for the photon, only the
longitudinal mode (which becomes apparent only at nonzero temperatures
on account of the Debye mass) contributes.}.  The constants $f_i$
assume the values $3/2$ for the scalars, fermions, and longitudinal
modes of the gauge bosons, and $1/2$ for transverse modes of the gauge
bosons~\cite{Wainwright:2011kj,Blinov:2015vma,Aoki:2021oez}.

Within this regularization framework, the choice of the
renormalization scale $\mu$ introduces an inherent uncertainty in the
determination of the critical temperature and other related
quantities~\cite{Chiang:2018gsn,Athron:2022jyi,Croon:2020cgk,Gould:2021oba}.
While, in the present analysis, we equate this scale to the mass of
the top quark\footnote{This is also the approximate geometric mean of
the electroweak symmetry breaking scale and the Higgs mass.}, the
scale dependence could be reduced by effecting a renormalization group
improvement of the effective potential (see,  e.g.,
Refs. ~\cite{Andreassen:2014eha,Andreassen:2014gha}). However, as
corrections on this account is expected to be small, we desist from
doing so.

The inclusion of the radiative corrections to the CW potential shifts
the position of the electroweak minimum and alters the physical Higgs
mass.  With the parameters of the potential
changing~\cite{Coleman:1973jx}, understandably, the tree-level
minimization conditions are no longer satisfied. To restore these
relations, a counter-term potential is introduced to cancel the
loop-induced shifts and ensure that the renormalized zero-temperature
potential reproduces the tree-level vacuum
configuration~\cite{Carrington:1991hz,Quiros:1999jp}. The counterterm
potential is written as
\begin{eqnarray}\label{eq:ctpot}
  V_{\rm ct}(h) &=& \frac{1}{2} \delta\mu_H^2 h^2
  + \frac{1}{4}\delta\lambda_H h^4,    
\end{eqnarray}
where the coefficients $\delta \mu_H^2$ and
$\delta\lambda_{H}$ can be derived by imposing the following
renormalization conditions at zero temperature\cite{Anisha:2022hgv}:
\begin{eqnarray}
\left. \frac{\partial \big(V_{\rm tree}+V_{\rm 1-loop}^{\rm CW}+V_{\rm ct}\big)}{\partial h} \right|_{h=v} = 0, \quad
\left. \frac{\partial^2 \big(V_{\rm tree}+V_{\rm 1-loop}^{\rm CW}+V_{\rm ct}\big)}{\partial h^2} \right|_{h=v} = m_h^2.
\end{eqnarray}
With the zero-temperature vacuum structure being preserved, the evolution of the electroweak phase transition is determined entirely by finite-temperature effects. The explicit expressions for $\delta\mu_H^2$ and $\delta\lambda_H$ in terms of the derivatives of the CW potential $V_{\rm 1-loop}^{\rm CW}(h)$ are provided in Appendix\,\ref{app:ct}.
 
The one-loop potential at $T\neq 0$ is given
by\cite{Dolan:1973qd,Weinberg:1974hy}:
\begin{equation}\label{eq:FT}
V^T_{\rm 1-loop}(h,T)=\frac{T^4}{2\pi^2}\left(\sum_{i=\rm Bosons }g_i J_B\left(\frac{m_i^2(h)}{T^2}\right) - \sum_{i=\rm Fermions }g_iJ_F\left(\frac{m_i^2(h)}{T^2}\right)\right),
\end{equation}
where the bosonic and fermionic thermal functions are
\begin{eqnarray}\label{eq:thermalF}
  J_{B/F}\Big(\frac{m^2(h)}{T^2}\Big)=\int_0^\infty  dx \, x^2 \log \left(1 \mp e^{-\sqrt{x^2+\frac{m^2(h)}{T^2}}}\right) .  
\end{eqnarray}
Clearly, for $m^2\gg T^2 $, the functions $J_{B, F}$ are suppressed
and the contributions to the effective potential from the heavy
particles in the loop would be negligible. On the other hand, for
$x\ll 1$ $(x\equiv\frac{m^2}{T^2})$,
these functions are well approximated
by~\cite{Quiros:1999jp} 
\begin{equation} \label{eq:expn}
  \barr{rcl}
  J_B(x) &= & \dis -\frac{\pi^4}{45} + \frac{\pi^2 x}{12}
  - \frac{\pi}{6} x^{3/2} - \frac{x^2}{32} \ln \frac{x}{a_b } + \cdots,
  \\ [2ex]
  J_F(x) &= & \dis \frac{7\pi^4 }{360} - \frac{\pi^2 x}{24}
     - \frac{x^2}{32} \ln \frac{x}{a_f} + \cdots, 
        \earr
\end{equation}
where $a_f = \pi^2\exp \left(3/2-2\gamma_E\right)$ and $a_b = 16 \pi^2\exp \left(3/2-2\gamma_E\right)$, with $\gamma_E \approx 0.577$ being the Euler-Mascheroni constant.  The generation of the effective $m^3(h)$ term in the Higgs potential---courtesy $J_B$---would turn out to be crucial for generating a false vacuum and, thereby, realizing a first-order phase transition.

While the finite temperature potential as described above is beset with infrared divergences, these are automatically cured on the inclusion of the ring (daisy) diagram contributions~\cite{Weinberg:1974hy,Carrington:1991hz} {\em viz.,}
\begin{equation}\label{eq:ring}
V_{\rm ring}(h ,T)=-\frac{T}{12\pi}g_i\left[m_i^3(h,T) -m_i^3(h)\right] \ .
\end{equation}
The thermal (Debye) masses are given
by~\cite{Weinberg:1974hy,Carrington:1991hz}.
\begin{equation}\label{eq:massT}
m_i^2(h) \rightarrow m_i^2(h,T) = m_i^2(h) +  \Pi_i(h,T) T^2 ,
\end{equation}
with the leading contributions emanating from the
field-dependent self-energy contributions $\Pi_i$ corresponding to the
daisy diagrams involving the top quark, the bosons ($W$, $Z$, the
Higgs itself and the Goldstones).  The result can be summarised
as~\cite{Postma:2020toi}
\begin{equation}\label{eq:HiggsTemp}
 \Pi_h(h,T) = \Pi_{\chi_i}(h,T) =  \frac{y_t^2}{4}  + \left(\frac{ g^2}{16}  +\frac{3 g^{\prime 2}}{16} \right) + \left(\frac{a_4}{2} + 4 a_6 h^2 \right) \, .
\end{equation}
It should be noted that the $1/\Lambda^2$ corrections to $y_t$, $g$, and $g^\prime$ from higher-dimensional operators give negligible contributions to the self-energy corrections $\Pi_i$, and thus to the phase transition strength parameter $v_c/T_c$. As for the resummed contributions from Majorana neutrinos in the loop, these are suppressed either by $h^2/\Lambda^2$ or by $m_{R_i}/\Lambda$ and, hence, are of little consequence.  On the other hand, the longitudinal polarization states of vector bosons get thermal corrected masses~\cite{Carrington:1991hz,Oikonomou:2024jms}:
\begin{align}
   & m_{W_L}^2(h) \rightarrow m_{W_L}^2(h,T)= m_{W_L}^2 (h)+ \Pi_{W_L}(T), \quad \Pi_{W_L}(T) = \frac{11}{6}g^2 T^2 .
\end{align}
At $T\neq 0$, there is mixing between the longitudinal modes of the
$Z$ boson and $\gamma$, and the temperature-dependent mass matrix is
given by~\cite{Carrington:1991hz,Oikonomou:2024jms}
\begin{equation}\label{eq:matrixZ}
   {\cal M}_{Z_L/\gamma_L}^2(h,T) =
    \begin{pmatrix}
    \frac{1}{4}g^2 h^2 + \frac{11}{6}g^2 T^2 & -\frac{1}{4}g g' h^2\\
    -\frac{1}{4}g g' h^2 & \frac{1}{4}g'^2 h^2 + \frac{11}{6}g'^2 T^2\\
    \end{pmatrix} \, ,
\end{equation}
and, thus,
\begin{equation}\label{Photon-thermalmass}
m^2_{Z_L/\gamma_L}(h,T) = 
(g^2 + g^{\prime 2}) \left(\frac{h^2}{8} + \frac{11}{12}  T^2\right)
  \pm \sqrt{\left(g^2 - g^{\prime 2} \right)^2 \left( \frac{h^2}{8}  + \frac{11T^2}{12}  \right)^2  + \frac{g^2 g^{\prime 2}}{16} h^4 }.
\end{equation}
Expectedly, at $T\neq 0$, the $\gamma_L$ is massive and contributes to the effective potential alongwith the remaining bosonic degrees of freedom, namely \(\{h, \chi_i, W_L, Z_L, W_T, Z_T\}\).

As discussed earlier, the resummation of thermal masses is essential for regulating infrared divergences and maintaining the perturbative consistency of the one-loop expansion~\cite{Senaha:2020mop,Kainulainen:2019kyp,Athron:2022jyi}.
This can be implemented using two equivalent but formally distinct prescriptions.  Ref.~\cite{Arnold:1992rz} advocates the
explicit addition of the ring correction term \(V_{\rm ring}(h,T)\) (see Eq.~\eqref{eq:ring}), accounting for the leading thermal
self-energy contributions from bosonic zero modes. Alternatively, the same effect is incorporated by replacing~\cite{Parwani:1991gq} the bosonic field-dependent masses \(m_i^2(h)\) with their thermally corrected forms \(m_i^2(h,T)\) directly in the one-loop CW and finite-temperature potentials, \(V^{\rm CW}_{1\text{-loop}}(h)\) and \(V^T_{1\text{-loop}}(h,T)\). We adopt the latter approach throughout
this work. The total one-loop effective potential, at non-zero temperatures can, thus, be written as 
\begin{equation}\label{eq:effpot}
  V_{\rm eff}(h,T) = V_0(h) + V^{\rm CW}_{1\text{-loop}}(h,T)
  + V^T_{1\text{-loop}}(h,T) + V_{\rm ct}(h),
\end{equation}
with the individual pieces \(V_0(h)\), \(V^{\rm
  CW}_{1\text{-loop}}(h,T)\), \(V^T_{1\text{-loop}}(h,T)\), and
\(V_{\rm ct}(h)\) being given by
Eqs.~\eqref{eq:pottree}, \eqref{eq:CW}, \eqref{eq:FT}, and
\eqref{eq:ctpot}, respectively. All terms in
the above equation possess an implicit temperature dependence arising
from the thermal evolution of the Higgs VEV, which
governs the shape of the effective potential and hence the nature of
the electroweak phase transition.

\subsection{First-order electroweak phase transition} \label{sec:FOPT}
We now investigate the model for the existence of a first-order
electroweak phase transition (FOEWPT), a possibility that is
non-existent within the SM. This would entail the following. At
sufficiently high temperatures, the only minimum of the potential
should be at $h = 0$ signifying the restoration of the electroweak
symmetry. As the universe cools down, a local minimum should start
developing at $h \neq 0 $ while $h = 0$ continues to be the global
minimum. At still lower temperatures, the two minima are of the same
height with a potential barrier between them, indicating the onset of
a first-order transition. As the universe cools further, the new local
minimum becomes the global minimum and finally the original minimum at
$h = 0$ is transformed into a local maximum.  The critical temperature
\(T_c\) is defined as the temperature at which the two minima are
degenerate, i.e.\ at \(h=0\) and \(h=v_c\), while the stationarity
condition determines the value of \(v_c\):
\begin{equation}
    \label{eq:veffeq}
    V_{\rm eff}(v_c,T_c)=V_{\rm eff}(0,T_c),
%
    \qquad \qquad
     \frac{\partial V_{\rm eff}(h,T_c)}{\partial h}\bigg|_{h=v_c} = 0.
\end{equation}

The strength of the FOPT is characterized by the ratio \(v_c/T_c\). A
strong transition requires $v_c \gtrsim T_c$. Within the Standard Model, this condition is satisfied only for a Higgs mass much lighter
than the observed value at the LHC. However, the dimension six
WCs $C_{\rm kin}$, $C_H$, and $C_{tH}$ significantly modify the
effective potential through their effects on the tree-level Higgs
potential and field-dependent masses. In particular, $C_H$, via
$\mathcal{O}_H$, introduces an explicit $h^6$ term, while $C_{\rm
  kin}$, through $\mathcal{O}_{HD}$ and $\mathcal{O}_{H\Box}$, induces
an effective $h^6$ contribution after canonical normalization; both
thereby alter the field-dependent scalar masses and reshape the
potential. In contrast, $C_{tH}$ affects the electroweak phase
transition indirectly by modifying the field-dependent top-quark mass,
contributing to the effective potential at the loop level.  The
$C_{H}$ contribution turns out to be the most important in
strengthening the phase transition. Its coefficient in the effective
potential being negative, $C_H > 0$ tends to destabilise the
potential. Indeed, obtaining a first-order phase transition requires
$C_H \lessapprox -0.5$ with the strength of the transition
increasing with $|C_H|$.

  \begin{figure}[h!]
    \centering
    \includegraphics[width=0.47\linewidth]{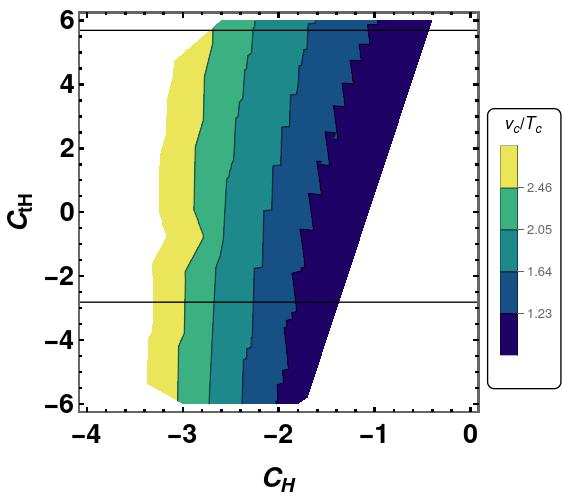}
    \includegraphics[width=0.47\linewidth]{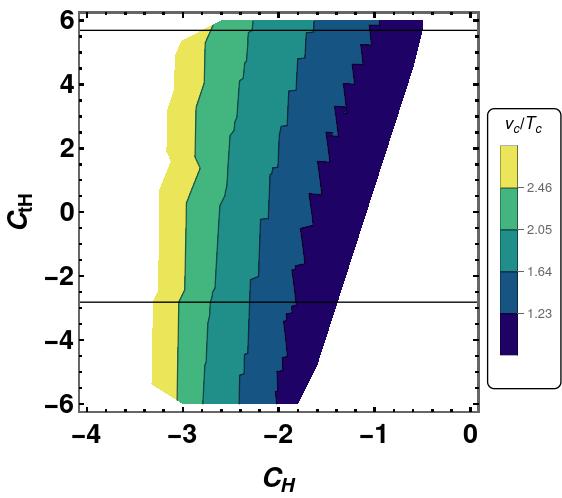}
   \caption{The ratio of the critical VEV to the critical temperature,
     \(v_c/T_c\), in the plane of \(C_H\) and \(C_{tH}\) for
     \(M_{i}=(200,200,300)\,\mathrm{GeV}\), \(C_{\rm kin}=0\), and
     $C_{NH_{33}}=0.001$. The left panel corresponds to
     \(c_{NH_{ii}}=1.0\), and the right panel to
     \(c_{NH_{ii}}=0.001\), assuming \(\Lambda=1\) TeV. The color codings refer to
     the value of \(v_c/T_c\). The horizontal black lines denote the constraints on $C_{tH}$ from the
     electroweak precision observables~\cite{Ellis:2020unq}.}
    \label{fig:smeftpt_1}
\end{figure}


In Fig.\,\ref{fig:smeftpt_1}, we show the estimated strength of the
FOEWPT in the plane of two specific SMEFT operators (assuming $C_{\rm
  kin}$ to be vanishing), with colored contours representing the
variation of \(v_c/T_c \in [1.0, 2.87]\).  In the
$\bigl(C_{H},\,C_{tH}\bigr)$ plane, the contours are nearly vertical,
indicating that the impact of $(C_{tH})$ on the transition is relatively small. The dominant behaviour is
controlled by $C_{H}$: increasingly negative values that deepen the
broken-phase minimum and substantially enhance the ratio
$v_{c}/T_{c}$. A cursory comparison of the two panels also shows that
$c_{NH}$ plays only a very subdominant role.

\begin{figure}[h!]
    \centering
    \includegraphics[width=0.47\linewidth]{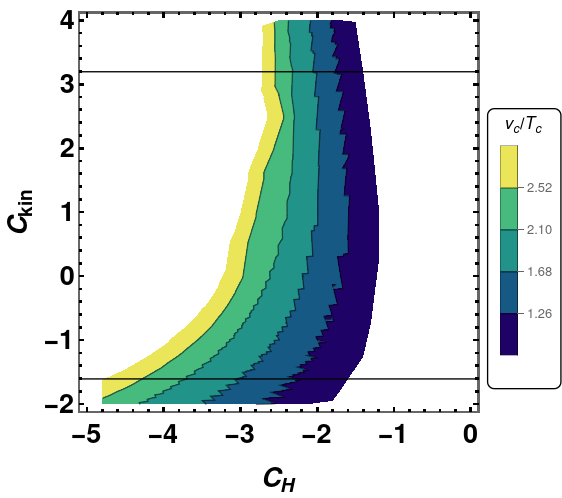}
    \includegraphics[width=0.47\linewidth]{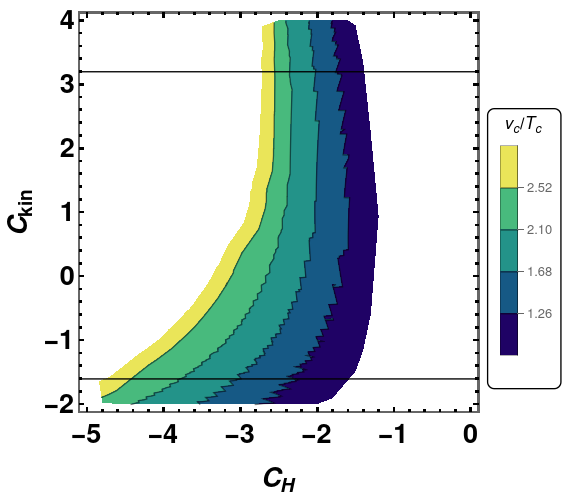}
   \caption{The ratio of the critical VEV to the critical temperature,
     \(v_c/T_c\), in the plane of \(C_H\) and \(C_{\rm kin}\) for
     \(M_{i}=(200,200,300)\,\mathrm{GeV}\), $C_{NH_{33}}=0.001$, and
     \(C_{tH}=0\). The left panel corresponds to \(c_{NH_{ii}}=1.0\),
     and the right panel to \(c_{NH_{ii}}=0.001\), assuming
     \(\Lambda=1\) TeV. The color codes refer to the value of \(v_c/T_c\). The
     horizontal black lines denote the constraints on $C_{\rm kin}$ from the electroweak
     precision observables~\cite{Ellis:2020unq}.}
    \label{fig:smeftpt_2}
\end{figure}

Switching on $C_{\rm kin}$ has a profound effect as it modifies the Higgs-field normalization
and shifts the effective quartic and thermal mass parameters, and its
effect becomes particularly significant once $C_{H}$ is
sufficiently negative. This is reflected, in
Fig.\,\ref{fig:smeftpt_2}, by the pronounced bending of the contours
at $C_{H}\lesssim -2$, indicating that even a
  moderate change in $C_{\rm kin}$ could significantly alter the strength of the
transition. In other words, while $C_{H}$ remains
the primary driver of a strong first-order transition, $C_{\rm kin}$
provides a second direction in parameter space that can further
enhance the transition strength for large-magnitude values of $C_{H}$.

 It should be noted that some of the SMEFT operators also contribute
 to electroweak precision observables as well as Higgs and top-quark
 measurements~\cite{Ellis:2020unq}. Of the variables of interest to
 us, $C_H$ remains relatively unconstrained, as it does not enter the
 relevant observables at the leading order. On the other hand, these
 observables are sensitive to $C_{tH}$ and $C_{\rm kin}$ and the
 corresponding bounds, derived from global fits, are shown as the
 solid black lines in the figure.

  Next, we examine the possible role of the \((\bar{N}^c N)(H^\dagger H)\) term in the phase transition dynamics. To simplify the discussion, we hold \(C_{tH}=0\) and \(C_{\text{kin}}=0\) as these have already been seen to be subdominant to $C_H$ in this regard. Furthermore, we consider two benchmark heavy-neutrino mass spectra in which \emph{two} of the three sterile states are nearly degenerate, a necessary condition for resonant leptogenesis, while the \emph{third} state is separated in mass.  Specifically, we study
\[
\text{Scenario A:}\quad M_i = (200,\,200,\,300)\,\mathrm{GeV}, \qquad
\text{Scenario B:}\quad M_i = (300,\,300,\,400)\,\mathrm{GeV}.
\]
For the neutrino--Higgs operator, we adopt two representative benchmark values
for the diagonal Wilson coefficients of the quasi-degenerate states,
\[
c_{NH_{ii}} = 0.001 \quad \text{and} \quad c_{NH_{ii}} = 1.0, \qquad i=1,2,
\]
corresponding to minimal and sizeable contributions to the effective
potential, respectively. In both scenarios, we fix the coefficient of the
non-degenerate state to \(c_{NH_{33}} = 0.001\).

\begin{figure}[h!]
    \centering
    \includegraphics[width=0.49\linewidth]{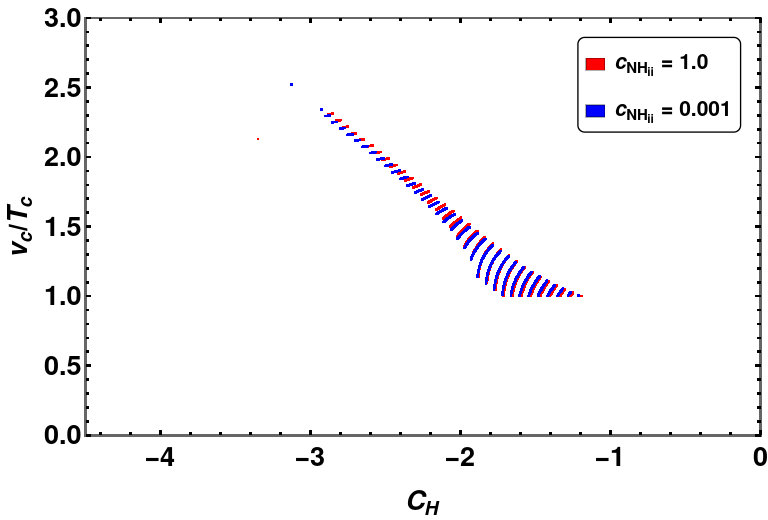}
    \includegraphics[width=0.49\linewidth]{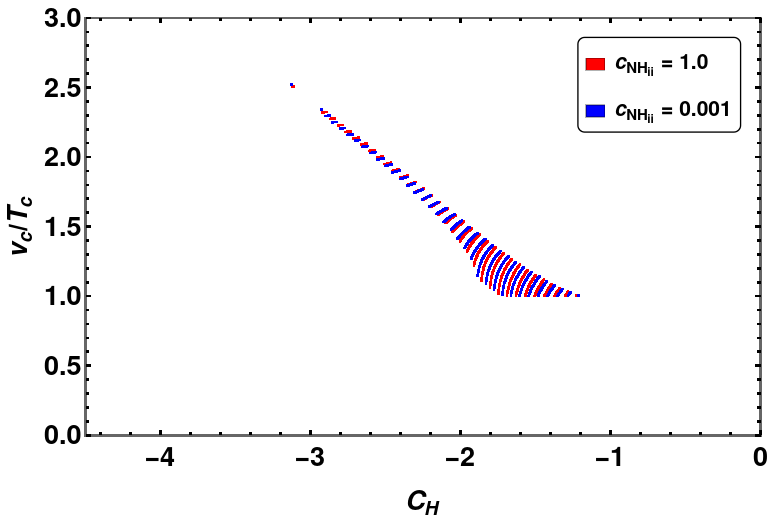}
    \caption{Phase transition strength \(v_c/T_c\) versus \(C_H\) for two heavy-neutrino mass scenarios: left for Scenario~A and right for Scenario~B. The additional Wilson coefficients are fixed at \(C_{\rm kin}=0\) and \(C_{tH}=0\), values consistent with LEP EWPO bounds. Regions with \(v_c/T_c\ge 1\) realize a strong first-order transition.}
    \label{fig:nonzeromass}
\end{figure}

Fig.\,\ref{fig:nonzeromass} displays the consequent transition
strength \(v_c/T_c\) as a function of \(C_H\). The \emph{left} panel
corresponds to Scenario~A, and the \emph{right} panel to
Scenario~B. In both cases, \(v_c/T_c\) increases with \(|C_H|\), so a
strong first-order EWPT (\(v_c/T_c>1\)) is realized once \(|C_H|\)
exceeds a modest threshold. The plots
confirm that \(C_H\) dominantly controls the FOEWPT strength, while
neutrino-induced corrections provide subleading modifications.

\subsection{Gravitational waves}\label{sec:gw}

A strong electroweak FOPT can generate a stochastic
  gravitational-wave background; we now compute the resulting spectrum
  and compare it with the sensitivities of present and future
  interferometers.

  In the GW study, we contrast Scenario~A and Scenario~B, which differ
in the mass spectrum of heavy neutrinos while sharing identical gauge
and portal couplings. This comparison isolates how the position of the
third neutrino relative to the electroweak scale imprints on the
thermodynamics of the EWPT and, consequently, on the stochastic GW
background.

To obtain quantitative predictions, we select representative benchmark parameter sets that (i) satisfy electroweak precision constraints and
(ii) realize a strong first-order EWPT. Specifically, we choose three benchmark points, BP1, BP2, and BP3, for the Wilson coefficients. The
corresponding values (with \(\Lambda = 1~\text{TeV}\)) are shown in Tab.\,\ref{tab:gwwp}. In the thermal history, symmetry is restored at high temperature; as the universe cools, the Higgs background develops a barrier between minima, the system supercools, and true-vacuum bubbles nucleate and percolate.


\begin{table}[H]
    \centering
    \begin{tabular}{|c|c|c|c|c|c|}
    \hline 
    Wilson coeffs.& BP1   & BP2 & BP3 \\
    \hline
         $C_H$ & - 3.2   & -3.5 & -4.0  \\
         \hline
        $C_{\rm kin}$ & 0.9  & 0  & -0.8  \\
        \hline
         $C_{tH}$ & 0  & 0.2 & 0.5  \\
        \hline
    \end{tabular}
    \caption{Benchmark values for scenario-I for WCs. We set $\Lambda$ to be 1 TeV.}
    \label{tab:gwwp}
\end{table}

As shown in our analysis, electroweak symmetry breaking proceeds through a first-order phase transition in the early Universe. At finite temperature, the effective potential develops two local minima separated by a barrier. As the Universe cools, it becomes temporarily trapped in the symmetric (false) vacuum before tunneling to the broken (true) vacuum, where the Higgs field acquires a nonzero vacuum expectation value. The transition proceeds through the nucleation and expansion of bubbles of the broken phase and provides a cosmological source of stochastic gravitational waves.

The dynamics of a first-order phase transition are conveniently characterized by a set of thermodynamic parameters evaluated at the nucleation temperature $T_N$. The most relevant quantities are the transition strength $\xi = v_c/T_c$, the latent heat parameter $\alpha_N$, and the inverse duration of the transition $\beta/H_N$. 

The strength of the phase transition is characterized by the latent heat parameter $\alpha_N$, defined as the ratio of the released vacuum energy to the radiation energy density at the nucleation temperature $T_N$,
\begin{equation}
\alpha_N \equiv \frac{\Delta \rho}{\rho_{\rm rad}} \, .
\end{equation}
The radiation energy density is given by
\begin{equation}
\rho_{\rm rad} = \frac{\pi^2\, g^\ast}{30}\, T_N^4 \, ,
\end{equation}
where $g^\ast$ denotes the effective number of relativistic degrees of freedom, which in the SM is $g^\ast_{\rm SM} = 106.75$~\cite{Husdal:2016haj}.
 The released vacuum energy $\Delta \rho$ is obtained from the finite-temperature effective potential as~\cite{Kamionkowski:1993fg, Kehayias:2009tn}
\begin{equation}
\Delta \rho =
\left[ V_{\rm eff}(\phi_0, T) - T \frac{dV_{\rm eff}(\phi_0, T)}{dT} \right]_{T = T_N}
-
\left[ V_{\rm eff}(\phi_N, T) - T \frac{dV_{\rm eff}(\phi_N, T)}{dT} \right]_{T = T_N},
\end{equation}
where $\phi_0$ and $\phi_N$ denote the Higgs field values in the symmetric and broken phases at $T_N$, respectively.

The inverse duration of the phase transition is characterized by the parameter $\beta/H_N$, defined as
\begin{equation}
\frac{\beta}{H_N}
=
T_N \frac{d}{dT}
\left( \frac{S_E(T)}{T} \right)\Bigg|_{T=T_N},
\end{equation}
where $H_N$ is the Hubble expansion rate at the nucleation temperature and Here $S_E(T)$ denotes the three-dimensional Euclidean action of the critical bubble at temperature $T$, which governs the thermal tunneling rate during the phase transition.

We compute these quantities numerically using \texttt{CosmoTransitions}~\cite{Wainwright:2011kj}, which determines the nucleation temperature by solving the condition $S_E(T_N)/T_N = 140$.
The resulting phase transition parameters for both Scenario~A and Scenario~B are summarized in Tab.\,\ref{tab:FOPT}. All benchmark points exhibit a strong first-order phase transition, with $\xi \gtrsim 2.6$, comfortably satisfying the sphaleron decoupling criterion. The critical temperatures lie in the range $T_c \simeq 80$--$85$~GeV, while bubble nucleation occurs at significantly lower temperatures, $T_N \simeq 50$--$60$~GeV. The Higgs field value $\phi_N(=v(T)|_{T_N})$ inside the broken phase at $T_N$ remains large, indicating a sizable potential barrier at nucleation.

A comparison of the two scenarios shows that Scenario~A typically yields larger values of $\alpha_N$ and smaller $\beta/H_N$, corresponding to a longer-lasting phase transition and a stronger gravitational wave signal. In contrast, Scenario~B exhibits larger $\beta/H_N$, indicating a faster transition and a shift of the gravitational wave spectrum toward higher frequencies. These differences play a central role in determining the spectral shape and detectability of the resulting stochastic gravitational wave background, as discussed in the following section.

\begin{table}[htbp]
\centering
\renewcommand{\arraystretch}{1.3}
\begin{tabular}{|c|c|c|c|c|c|c|}
\hline
\multirow{2}{*}{PT parameters}
 & \multicolumn{3}{c|}{Scenario A}
 & \multicolumn{3}{c|}{Scenario B} \\
\cline{2-7}
 & BP1 & BP2 & BP3 & BP1 & BP2 & BP3 \\
\hline
$v_c$ [GeV]
 & 227 & 225 & 224 & 227 & 225 & 224 \\
\hline
$T_c$ [GeV]
 & 82 & 83 & 84 & 83 & 84 & 85 \\
\hline
$\xi = v_c/T_c$
 & 2.77 & 2.71 & 2.67 & 2.73  & 2.68 & 2.64 \\
 \hline
$T_N$ [GeV]
 & 45 & 53 & $57$ & 48 & 54 & 59 \\
\hline
$v_N$ [GeV]
 & 245 & 244 & 242 & 245 & 243 & 242 \\
\hline
$\alpha_N$
 & 0.32 & $0.21$ & $0.17$ & 0.27 & 0.19 & $0.16$ \\
\hline
$\beta/H_N$
 & 113 & 331 & $295$  & 133 & 186 & $266$  \\
\hline
\end{tabular}
\caption{Phase transition parameters for $C_{NH_{ii}}=1.0\, (i=1,2)$.}
\label{tab:FOPT}
\end{table}

Using the phase transition parameters obtained above, we now evaluate the stochastic gravitational wave signal generated during the transition. The spectrum is determined by the dynamics of bubble nucleation and expansion, as well as by the subsequent interactions between the expanding bubble walls and the surrounding plasma.

The GW energy density spectrum consists of three primary contributions: bubble wall collisions, sound waves, and magneto-hydrodynamic turbulence (MHD). These components collectively determine the total energy spectrum, which can be approximated as the sum of all three:
\begin{equation}\label{eq:GWTotal}
\Omega_{\rm GW}h^2 \approx \Omega_{\rm col} h^2 + \Omega_{\rm sw} h^2 + \Omega_{\rm tur} h^2,
\end{equation}
where the terms on the right-hand side represent the contributions from bubble wall collisions, sound waves, and MHD turbulence, respectively \cite{Caprini:2015zlo}. Here, \( h = H_0/(100 \, \text{km} \cdot \text{s}^{-1} \cdot \text{Mpc}^{-1}) \), where \( H_0 \) is the present-day Hubble constant \cite{DES:2017txv}.

The energy density contribution from bubble wall collisions can be derived using the envelope approximation \cite{Jinno:2016vai} and is expressed as a function of frequency \( f \):
\begin{equation}\label{eq:GWcoldetails}
\Omega_{\rm col} h^2 = 1.67 \times 10^{-5} {\left(\frac{\beta }{H_N} \right)^{-2}} \left( \frac{\kappa_c \alpha_N}{1 + \alpha_N} \right)^2 \left(  \frac{100}{g^{\ast}} \right)^{1/3} \left( \frac{0.11 v^3_w}{0.42 + v^2_w} \right) \frac{3.8 \left( f/f_{\rm col} \right)^{2.8}}{1 + 2.8 \left( f/f_{\rm col} \right)^{3.8}},
\end{equation}
where \( v_w \) is the bubble wall velocity, \( g_* \) is the effective degrees of freedom and for SM it is $106.75$~\cite{Husdal:2016haj}, and \( \kappa_c \) is the efficiency factor for bubble collisions, defined as:
\begin{equation}\label{eq:kcfac}
\kappa_c = \frac{0.715 \alpha_N + \frac{4}{27} \sqrt{\frac{3 \alpha_N}{2}}}{1 + 0.715 \alpha_N}.
\end{equation}
The redshifted peak frequency \( f_{\rm col} \) is given by:
\begin{equation}\label{eq:PF1}
f_{\rm col} = 16.5 \times 10^{-6} \left( \frac{f_{\ast}}{\beta} \right) \left( \frac{\beta}{H_N} \right) \left( \frac{T_N}{100 \, {\rm GeV}} \right) \left( \frac{g^{\ast}}{100} \right)^{1/6} \, {\rm Hz},
\end{equation}
where the fitting function \( f_{\ast}/\beta \) is defined as \cite{Jinno:2016vai}:
\begin{equation}\label{eq:fastbybetadetails}
\frac{f_{\ast}}{\beta} = \frac{0.62}{1.8 - 0.1 v_w + v^2_w}.
\end{equation}
Assuming \( v_w = 1 \), as expanding bubbles achieve relativistic terminal velocities, simplifies the calculation.

The energy density contribution from sound waves can be expressed as \cite{Hindmarsh:2013xza,Hindmarsh:2016lnk,Hindmarsh:2017gnf}:
\begin{equation}\label{eq:GWswpart}
\Omega_{\rm sw} h^2 = 2.65 \times 10^{-6}\; \Upsilon(\tau_{\rm sw}) \left(  \frac{\beta}{H_N} \right)^{-1} v_w \left( \frac{\kappa_{\rm sw} \alpha_N}{1 + \alpha_N} \right)^2 \left( \frac{100}{g^{\ast}} \right)^{1/3} \left( \frac{f}{f_{\rm sw}} \right)^3 \left[ \frac{7}{4 + 3 \left( f/f_{\rm sw} \right)^2} \right]^{7/2},
\end{equation}
where \( \kappa_{\rm sw} \) is the efficiency factor indicating the fraction of latent heat converted into bulk plasma motion~\cite{Kamionkowski:1993fg}
\begin{equation}\label{eq:kappasw}
\kappa_{\rm sw} =\frac{\sqrt{\alpha_N}}{0.135+\sqrt{0.98+\alpha_N}}
\end{equation}
The peak frequency for sound waves is given by:
\begin{equation}\label{eq:PF2}
f_{\rm sw} = 1.9 \times 10^{-5} \left( \frac{1}{v_w} \right) \left( \frac{\beta}{H_N} \right) \left( \frac{T_N}{100 \, {\rm GeV}} \right) \left( \frac{g^{\ast}}{100} \right)^{1/6} \, {\rm Hz}.
\end{equation}
The factor \( \Upsilon(\tau_{\rm sw}) \), which accounts for the finite lifetime of sound waves, is defined as:
\begin{equation}\label{eq:swtimepart}
\Upsilon(\tau_{\rm sw}) = 1 - \frac{1}{\sqrt{1+2 \tau_{\rm sw} H_{\ast}}},
\end{equation}
where \( \tau_{\rm sw} \) is the lifetime of the sound wave. Following Ref.~\cite{Hindmarsh:2017gnf}, we write $\tau_{\rm sw}\approx R_N/\overline{U}_f$, where the mean bubble separation $R_N = (8\pi)^{1/3}v_w \beta_N^{-1}$ and the root-mean-squared fluid velocity $\overline{U}_f=\sqrt{3\kappa_{\rm sw}\alpha/4}$. 

Finally, the contribution from MHD turbulence, caused by the complete ionization of the plasma \cite{Caprini:2009yp}, is expressed as:
\begin{equation}\label{eq:GWturpart}
\Omega_{\rm tur} h^2 = 3.35 \times 10^{-4} \left( \frac{\beta}{H_N} \right)^{-1} v_w \left( \frac{\kappa_{\rm tur} \alpha_N}{1 + \alpha_N} \right)^{3/2} \left( \frac{100}{g^{\ast}}\right)^{1/3} \left[ \frac{\left( f/f_{\rm tur} \right)^3}{\left[ 1 + \left( f/f_{\rm tur} \right) \right]^{11/3} \left( 1 + \frac{8 \pi f}{h_{\ast}} \right)} \right],
\end{equation}
where \( h_{\ast} = 16.5\times 10^{-6} \left( \frac{T_N}{100 \, {\rm GeV}} \right) \left( \frac{g^{\ast}}{100} \right)^{1/6} \, {\rm Hz} \), the inverse Hubble time during GW production, redshifted to today. The peak frequency \( f_{\rm tur} \) is:
\begin{equation}\label{eq:PF3}
f_{\rm tur} = 2.7 \times 10^{-5} \frac{1}{v_w} \left( \frac{\beta}{H_N} \right) \left( \frac{T_N}{100 \, {\rm GeV}} \right) \left( \frac{g^{\ast}}{100} \right)^{1/6} \, {\rm Hz.}
\end{equation}
The turbulence efficiency factor \( \kappa_{\rm tur} \) is defined as
\( \kappa_{\rm tur} = \epsilon \kappa_{\rm sw} \), where \( \epsilon
\) represents the fraction of bulk motion converted into
turbulence. Previous studies suggest \( \kappa_{\rm tur} \approx 0.1
\kappa_{\rm sw} \), a value adopted here for numerical calculations.

\begin{figure}
    \centering
    \includegraphics[width=0.48\linewidth]{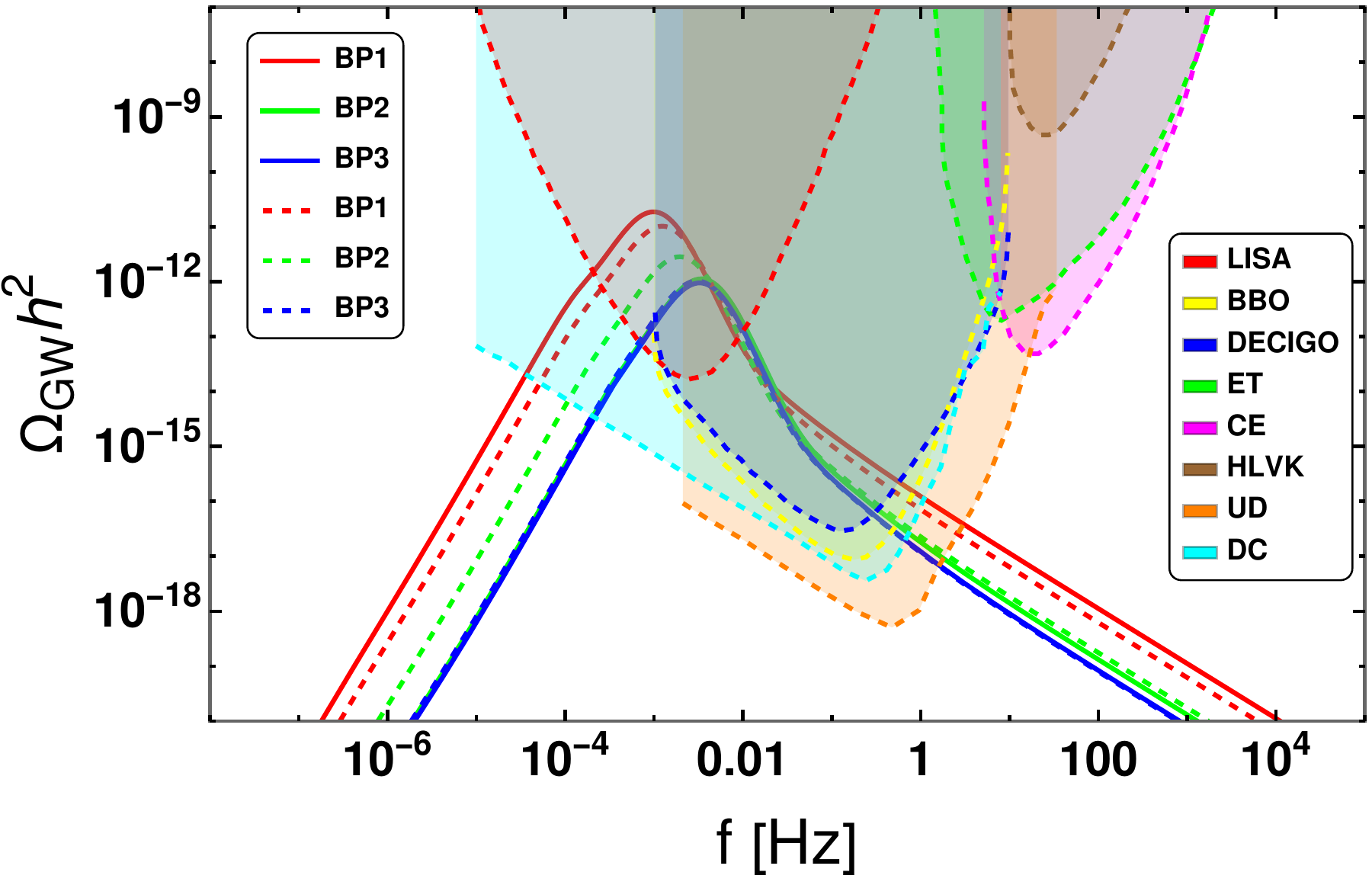}
    \includegraphics[width=0.48\linewidth]{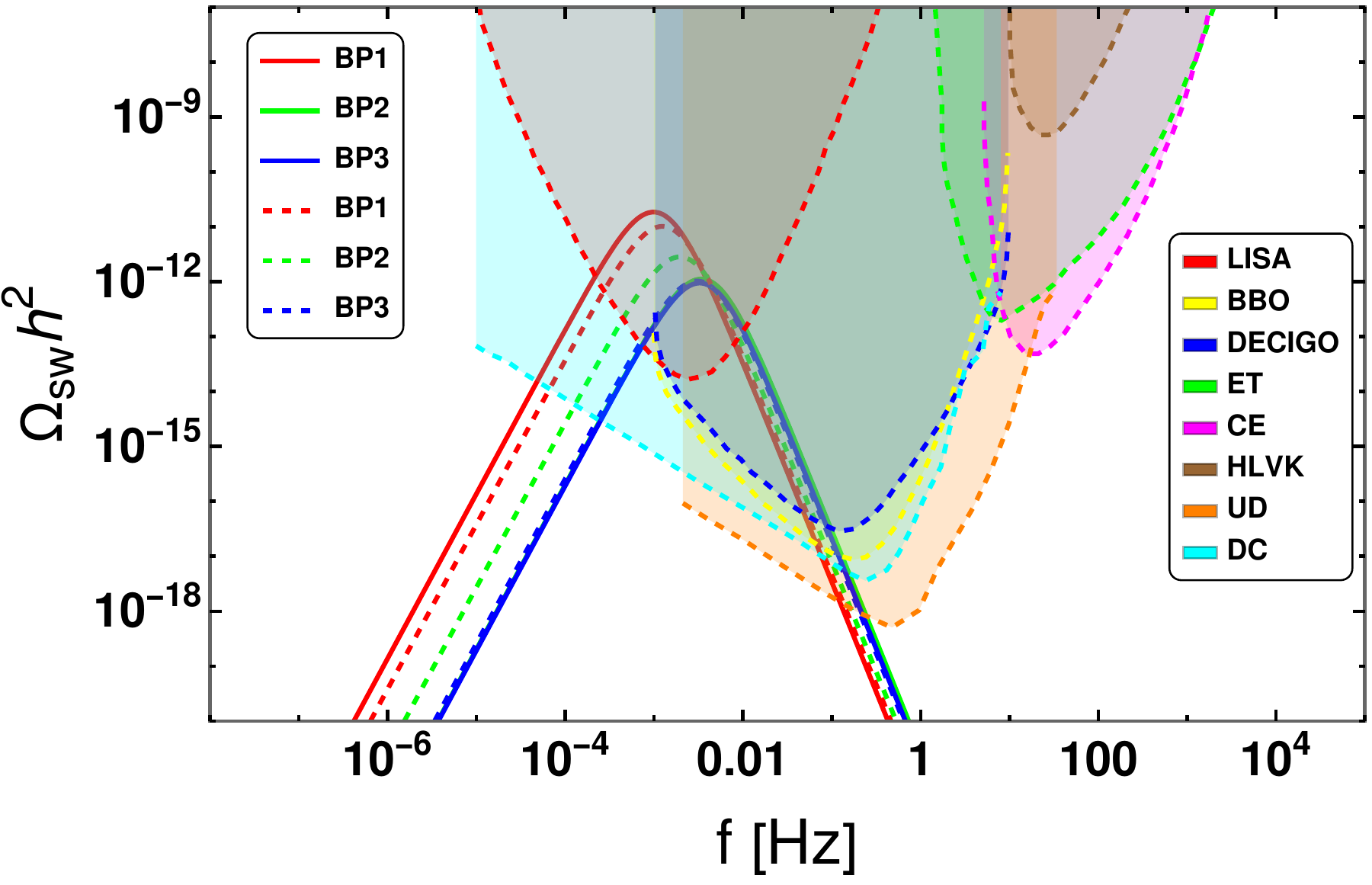}
    \includegraphics[width=0.48\linewidth]{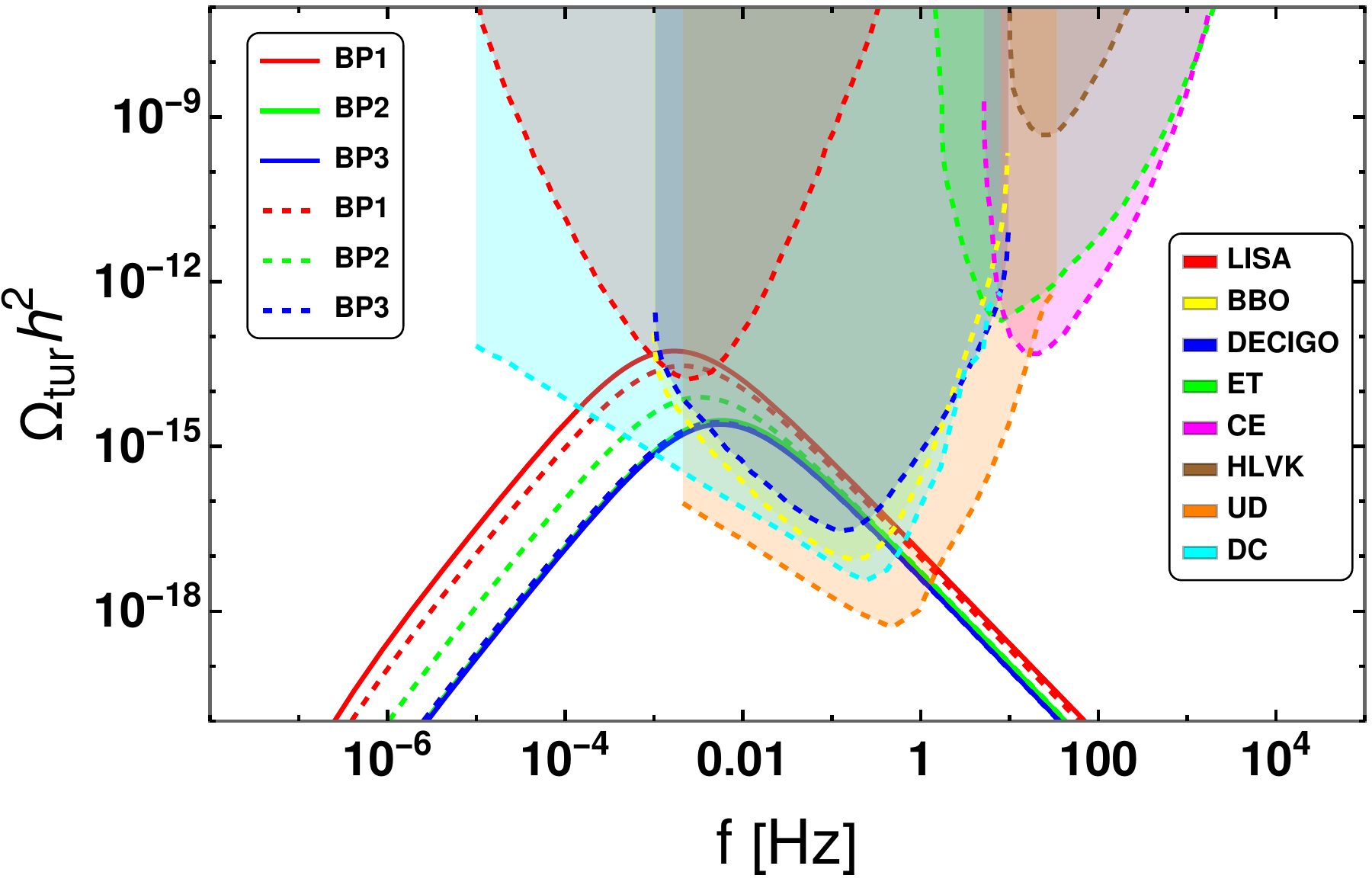}
    \includegraphics[width=0.48\linewidth]{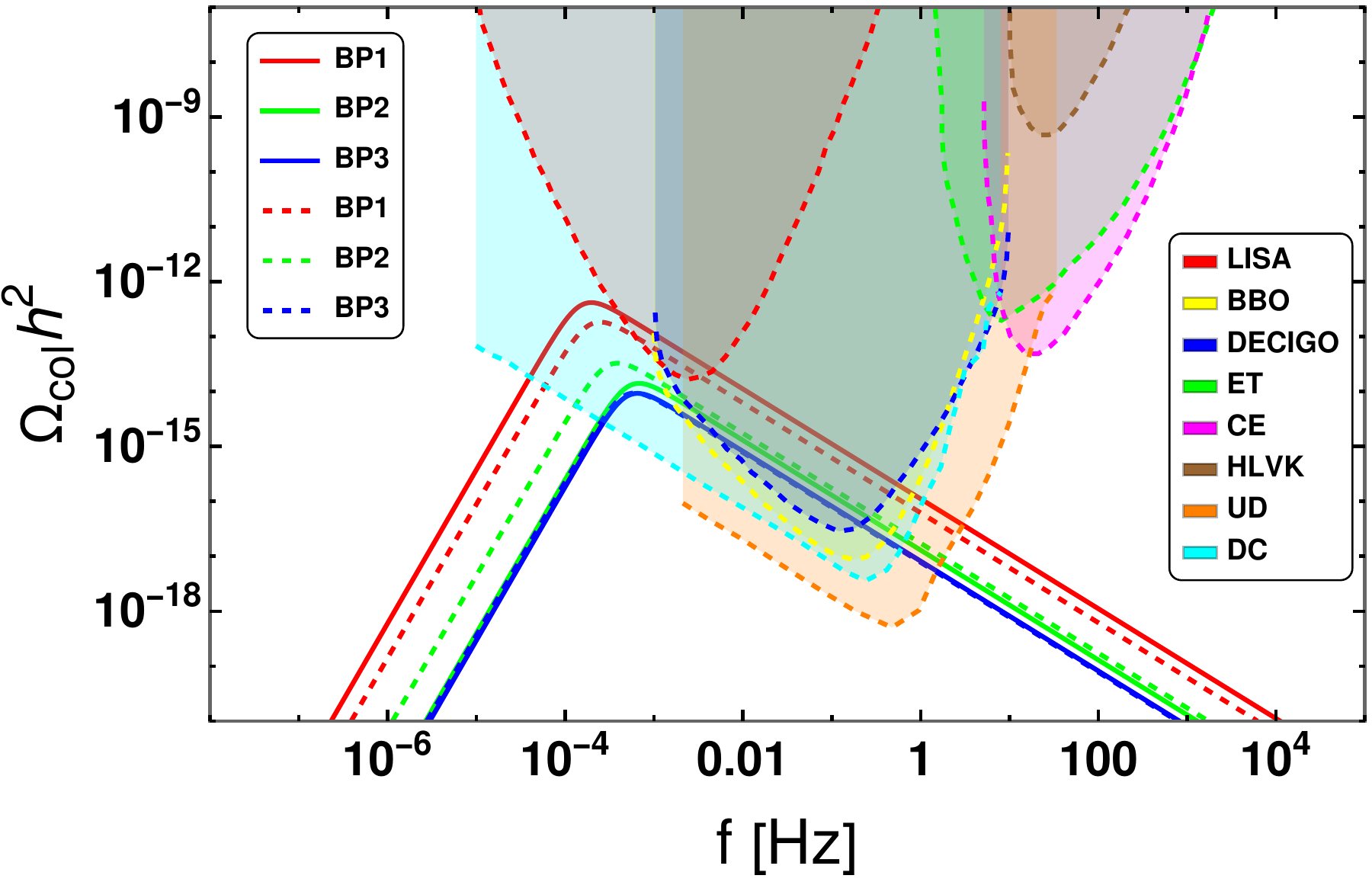}
    \caption{Variation of gravitational wave amplitudes with frequency. Top Left: resultant amplitudes, Top Right: only for sound waves. Bottom Left: only for turbulence, Bottom Right: only for bubble collisions.  Solid and dashed curves for $M_{i}=(200,200,300)$ GeV (scenario A) and $M_{i}=(300,300,400)$ GeV (scenario B), respectively. For both scenarios, we have taken $c_{NH_{ii}}=1.0$. The shaded regions show the projected sensitivity curves of future GW interferometers, including LISA~\cite{LISA:2017pwj}, BBO~\cite{Yagi:2011wg}, DECIGO~\cite{Nakayama:2009ce}, the Einstein Telescope (ET)~\cite{Punturo:2010zz}, Cosmic Explorer (CE)~\cite{Reitze:2019iox}, the HLVK network (LIGO Hanford–Livingston, Virgo, and KAGRA)~\cite{LIGOScientific:2014pky,VIRGO:2014yos,KAGRA:2018plz}, ultimate DECIGO (UD), and DECIGO-correlation (DC)~\cite{Nakayama:2009ce}, as indicated in the inset legend.}
    \label{fig:gwplots}
\end{figure}

Fig.\,\ref{fig:gwplots} shows the stochastic gravitational wave
spectrum produced by the electroweak phase transition as a function of
frequency, with the individual contributions from bubble wall
collisions, sound waves, and magneto-hydrodynamic turbulence displayed
separately. The predicted spectra are compared with the projected
sensitivities of current and future gravitational wave detectors,
including LISA~\cite{LISA:2017pwj}, BBO~\cite{Yagi:2011wg}, Einstein
Telescope (ET)~\cite{Punturo:2010zz}, Cosmic Explorer
(CE)~\cite{Reitze:2019iox},
HLVK~\cite{LIGOScientific:2014pky,VIRGO:2014yos,KAGRA:2018plz},
DECIGO, ultimate-DECIGO, and DECIGO with
correlation~\cite{Nakayama:2009ce}. In all benchmark cases, the
spectrum is dominated by the sound-wave contribution, while turbulence
yields a subleading high-frequency tail and bubble wall collisions
contribute negligibly, as expected for non-runaway bubble walls.

For Scenario~A, corresponding to $M_{i}=(200,200,300)$~GeV (solid
curves), the peak of the spectrum lies at frequencies of a few
millihertz, and the overall amplitude is relatively large, placing the
signal well within the projected sensitivity of LISA, BBO, and
DECIGO. In Scenario~B, with $M_i=(300,300,400)$~GeV (dashed curves),
the slightly higher nucleation temperature, together with a
substantially larger $\beta/H_N$ shifts the peak toward higher
frequencies, in the range of tens of millihertz to
$\mathcal{O}(0.1)$~Hz, while suppressing the amplitude approximately
as $(\beta/H_N)^{-1}$. Consequently, the signal in Scenario~B is
optimally matched to BBO and DECIGO, and only marginally accessible to
LISA. Across all panels, the total spectrum closely follows the
sound-wave contribution, confirming its dominant role, whereas
turbulence mainly affects the high-frequency tail and bubble
collisions remain subleading. Overall, the projected detector
sensitivities indicate that both scenarios are testable, with
Scenario~A, favoring lower-frequency observatories and Scenario~B
favoring higher-frequency ones.

The strength parameter $\alpha_N$ encodes the qualitative features of the phase transition. As shown in Tab.\,\ref{tab:FOPT}, Scenario~A typically yields larger values of $\alpha_N$ than Scenario~B. Since $\alpha_N$ measures the vacuum energy released at the nucleation temperature relative to the radiation energy density, larger values correspond to a stronger FOPT and hence a stronger GW signal. This can be understood as follows: in Scenario~A, the heavy Majorana states that dominate the finite-temperature effective potential lie closer to the electroweak scale, leading to more pronounced thermal effects and a stronger transition, whereas in Scenario~B they are more decoupled, resulting in a smaller $\alpha_N$.

\section{Neutrino masses}
\label{sec:neumass}

The singlet fermions acquire Majorana masses, both large bare terms,
as well as suppressed contributions from the dimension-5 operators
after EWSB.  In addition, the $Z_2$-even ones are involved in Yukawa
terms, leading to their mixings with the SM neutrinos. Postponing the
discussion of the $Z_2$-odd field, the relevant terms in the
Lagrangian are
\begin{equation}
\mathcal{L}_N \supset
-\frac{1}{2}\,m_{R_{ij}}\,\overline{N_i^{c}}N_j
-\mathcal{Y}_{D\,\alpha i}\,\overline{L_\alpha} \tilde H N_i
-\frac{c_{NH_{ij}}}{2\Lambda}(H^\dagger H)\,\overline{N_i^{c}}N_j
+\text{h.c.},
\end{equation}
where $L_\alpha$ denotes the lepton doublet with $\alpha=e,\mu,\tau$,
and $N_i$ are the gauge-singlet fermions. With $N_3$ being largely decoupled from the rest,  neutrino-mass generation as well as leptogenesis is driven only by the two heavy states $N_{1,2}$.

After electroweak symmetry breaking, the higher-dimensional operator induces the correction
\begin{equation}
m_N \equiv m_R + \frac{c_{NH}v^2}{2\Lambda},
\end{equation}
and the heavy-light mixing mass matrix in the $(\nu_L\, N^c)^T$ basis takes the form
\begin{equation}
\mathcal{M}_\nu=
\begin{pmatrix}
0 & m_D \\
m_D^T & m_N
\end{pmatrix},
\qquad
m_D=\frac{v}{\sqrt{2}}\,\mathcal{Y}_D,
\end{equation}
where $m_D$ is a $3\times 2$ matrix and $m_N$ is a $2\times 2$ complex
symmetric matrix. The full matrix
$\mathcal{M}_\nu$ is therefore complex symmetric of dimension
$5\times5$. Without any loss
  of generality, the matrix $m_N$ may be taken to be a real diagonal
  one. More specifically, however, we assume it to be (nearly)
  degenerate, a condition required to efficiently drive leptogenesis.
  It should be noted that if this degeneracy is to be occasioned by
  a symmetry argument connecting $N_{1,2}$, it would be equally applicable to
  both $m_N$ and $c_{NH}$. In other words, we have
\[
m_R  \approx \mathrm{diag}(m_{R_1},m_{R_2}),
\]
with $m_{R_1}\simeq m_{R_2}$ at the electroweak scale. Indeed, we
would consider an exact equality, assuming it to be the consequence of
an unspecified symmetry.

The Dirac Yukawa matrix is constructed using the Casas–Ibarra parameterization~\cite{Casas:2001sr} adapted to the two heavy state case, namely
\begin{equation}
\mathcal{Y}_D=-i\frac{\sqrt{2}}{v}U\,\sqrt{D_{m_\nu}}\,R\,\sqrt{D_{m_N}}\,,
\end{equation}
where $U$ is the Pontecorvo–Maki–Nakagawa–Sakata (PMNS)
matrix~\cite{ParticleDataGroup:2020ssz} with the best-fit parameters
taken from~\cite{Esteban:2020cvm}, and $R$ is a complex orthogonal
matrix of dimension $3\times2$~\cite{Casas:2001sr}. $D_{m_\nu}$ and
$D_{m_N}$ denote the masses of the light and heavy neutrinos
respectively, {\em viz.}, $D_{m_\nu} = \mathrm{diag}( 0, m_{\nu_2},
m_{\nu_3})$ and $D_{m_N} = \mathrm{diag}(M_1, M_2)$, corresponding to
the normal hierarchy for light neutrino masses. Since only two heavy
states participate in the seesaw mechanism, one active neutrino
remains massless. The light-neutrino masses and mixing angles are
fixed to the oscillation data~\cite{Esteban:2020cvm}, and the
cosmological bound~\cite{Planck:2018vyg} of $\sum_i
m_{\nu_i}<0.12\ \mathrm{eV}$ is trivially satisfied.

\subsection{Constraints from CLFV processes}\label{sec:clfv}
The heavy neutrinos can mediate charged lepton flavor-violating
processes such as $\ell_i\rightarrow\ell_j \gamma$, $\ell_i\rightarrow
\ell_j \ell_k \ell_l$, and $\mu $ to $e$ conversion. Experiments put a
stringent limit on the branching ratios for these
processes~\cite{Belle:2007qih,Crivellin:2020klg,Capdevila:2020rrl},
namely
\begin{eqnarray}
   &BR& (\mu \to e \gamma) \leq 4.2 \times 10^{-13},\quad BR (\tau \to \mu \gamma) \leq 4.5 \times 10^{-8}, \quad BR (\tau \to e \gamma) \leq 1.2 \times 10^{-9}, \nonumber \\
    & BR& (\mu \to 3e) \leq 1.0 \times 10^{-12},\quad BR (\tau \to 3\mu ) \leq 2.1 \times 10^{-8}, \quad BR (\tau \to 3e ) \leq 2.7 \times 10^{-8}, \nonumber \\
    & BR& (\tau \to \mu e e) \leq 1.5 \times 10^{-8},\quad BR (\tau \to  e \mu \mu ) \leq 1.7 \times 10^{-8}, \, CR(\mu N \to e N)\leq 7.0 \times 10^{-13} \nonumber
\end{eqnarray}
with the last named being the conversion rate. Contributions in the
present case are similar to those in type-I seesaw models, and the
analytic expressions thereof are relegated to
Appendix~\ref{app:lfvexp}.

For the numerical analysis to delineate the parameter space available
to the model, we start with the neutrino oscillation data.  To be
specific, we restrict ourselves to normal mass hierarchy, which then
fixes $m_{\nu_{2,3}}$ in $D_{m_\nu}$ within experimental error
bars. Similarly, the PMNS matrix $U$ too is determined (within the
attendant uncertainties, the latter being substantial for the
CP-violating phase).  Within the two-heavy-neutrinos scenario, the
Casas--Ibarra matrix $R$ is a complex orthogonal matrix of dimension
$2\times3$, which can be parametrized in terms of a single complex
angle,
\[
\omega = x + i y .
\]
Consequently, the Yukawa matrix $\mathcal{Y}_D$ depends only on the
two real parameters $(x,y)$, which, then, also parametrise the rates
of charged lepton-flavour-violating processes (CLFV). A numerical scan
is performed imposing both neutrino oscillation data and the CLFV
constraints.  Since the Yukawa couplings are invariant under $x \to x
+ \pi$, we restrict
\[
0 \le x \le \pi \ , \qquad |y| < 3.
\]
as larger $|y|$ leads to excessively large Yukawa couplings and are
excluded by CLFV constraints.

\section{Leptogenesis} \label{sec:leptogenesis}

Given that the model naturally admits lepton-number violation, a
non-zero lepton asymmetry could be generated in this sector through
the interference between the tree-level and one-loop diagrams
contributing to the heavy neutrino decay processes. This may,
subsequently, be transmuted to a baryon asymmetry through the
sphaleron processes at the electroweak scale.

A key ingredient is the relative difference in the decay widths of the
heavy neutrinos into leptons and anti-leptons, {\em viz.}
\begin{equation}\label{eq:CP}
  \epsilon_{i}=\frac{\Gamma(N_i \to L H)-\Gamma(N_i \to \bar{L} {H}^\dagger)}
          {\Gamma(N_i \to L H)+\Gamma(N_i \to \bar{L} H^\dagger)}.
\end{equation}
In Fig.\,\ref{fig:diags}, we illustrate the relevant decay diagrams.
It should be noted that we are interested in physics above the
symmetry breaking scale, and hence $L_\alpha$ and $H$ are the complete
doublet fields.  For each external state $L_\alpha$, the contributions
of individual $L_\beta$ loops, of course, need to be summed
over. Furthermore, $\epsilon_i$ is inclusive in the sense that all
final state lepton flavours are implicitly summed over\footnote{The
differences in the lepton masses are of little consequence.}.  For
nearly degenerate $N_{1,2}$, the CP asymmetry is dominated by the
self-energy diagram, with the vertex correction contributions being
numerically subleading and essentially negligible in the parameter
region relevant for our analysis. The tree-level decay widths of the
\( N_i \) at temperature $T$ are given
by~\cite{Pilaftsis:2013nqa,Giudice:2003jh}
%
%
%
\begin{figure}[!b]
\centering
\begin{tikzpicture}[scale=1.0]

\begin{scope}
\node at (-0.5,0) {$N_i$};
\draw (-0.2,0) -- (1,0);
\draw[dashed] (1,0) -- (2,0.8);
\draw (1,0) -- (2,-0.8);        
\node at (2.2,0.9) {$H$};
\node at (2.2,-0.9) {$L_\alpha$};
\end{scope}
\begin{scope}[xshift=3.5cm]
\node at (-0.5,0) {$N_i$};
\draw (-0.2,0) -- (0.8,0);
\draw (0.8,0) arc (180:0:0.6);        
\draw[dashed] (2.0,0) arc (0:-180:0.6); 
\node at (1.4,0.9) {$L_\beta$};
\node at (1.4,-0.9) {$H$};
\node at (2.5,0.3) {$N_j$};
\draw (2.0,0) -- (3.0,0);
\draw[dashed] (3.0,0) -- (4.0,0.8); 
\draw (3.0,0) -- (4.0,-0.8);       
\node at (4.2,0.9) {$H$};
\node at (4.2,-0.9) {$L_\alpha$};
\end{scope}

\begin{scope}[xshift=9cm]
\node at (-0.5,0) {$N_i$};
\draw (-0.2,0) -- (1,0);
\draw (1,0) -- (2,0.8);          
\draw[dashed] (1,0) -- (2,-0.8); 
\draw (2,0.8) -- (2,-0.8);       
\node at (1.3,0.8) {$L_\beta$};
\node at (1.3,-0.8) {$H$};
\node at (2.3,0) {$N_j$};
\draw[dashed] (2,0.8) -- (3,0.8);
\draw (2,-0.8) -- (3,-0.8);       
\node at (3.2,0.8) {$H$};
\node at (3.3,-0.8) {$L_\alpha$};
\end{scope}

\end{tikzpicture}
\caption{Diagrams contributing to asymmetric decay of $N_i$.\label{fig:diags}}
\end{figure}
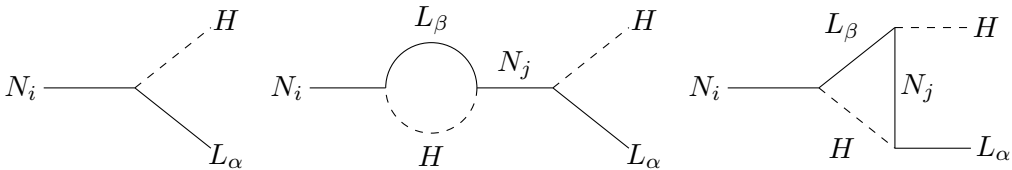
\begin{eqnarray} 
\Gamma^T_{N_i}&=&\frac{(\mathcal{Y}_D^\dagger \mathcal{Y}_D)_{ii} M_{i}}{8\pi} \lambda^{1/2}(1,a_H,a_L) (1-a_H+a_L) 
= x_p \Gamma^0_{N_i},
\label{N_width}
\end{eqnarray}
where $\Gamma^0_{N_i}$ corresponds to the vacuum decay width of $N_i$
\textit{i.e.,} decay width at zero temperature. Here, the factor $x_p
=\lambda^{1/2}(1,a_H,a_L)(1-a_H+a_L)$ accounts for the modified
two-body phase space when the Higgs and lepton doublets acquire
thermal masses, and the
K\"all\'en function is defined as
$\lambda(a,b,c)=a^{2}+b^{2}+c^{2}-2ab-2ac-2bc$.
The dimensionless parameters are defined by 
$a_{H/L}=\frac{m^2_{H/L}(T)}{M_i^2}$, 
where $M_i$ denotes the heavy neutrino mass\footnote{Thermal corrections to Majorana neutrino
masses scale as $\mathcal{Y}_D^2$ and are typically small in the
relevant parameter space; hence, they are neglected in the absolute
masses but retained when considering mass splittings.}, and $m_H^2(T)(=c_H T^2)$ and $m_L^2(T)(=c_L T^2)$ are the entirely temperature-dependent mass-squared of the Higgs doublet (same as $\Pi_h T^2$ given in Eq.~\eqref{eq:HiggsTemp}) and the lepton doublet given by~\cite{Giudice:2003jh,Weldon:1982bn}
\begin{eqnarray}
   m_L^2(T)= \left(\frac{3g^2}{32} + \frac{g^{\prime 2}}{32} \right) T^2\equiv c_L T^2.
\end{eqnarray} 
In the considered temperature range, the temperature induced mass of
the Higgs doublet is comparable to the mass of the decaying particle
and therefore cannot be neglected. On the other hand, at sufficiently
high temperatures, the Higgs doublet acquires a thermal mass
correction that exceeds the corresponding mass correction of the
lepton doublet. As for the numerator in the asymmetries
$\epsilon_i$, the self-energy diagram contribution dominates over the
vertex correction in the quasi-degenerate regime. If one neglects the
motion of the decaying particle (\textit{i.e.,} rest in the thermal
bath.), the leading contributions to the CP-asymmetry
is~\cite{Pilaftsis:2013nqa,Covi:1997dr,Giudice:2003jh}
\begin{eqnarray}\label{eq:cp_asym}
\epsilon_{ij}(T)&\approx& \dis
\frac{\text{Im}[\mathcal{Y}_D^\dagger \mathcal{Y}_D]_{ij}^2}{(\mathcal{Y}_D^\dagger \mathcal{Y}_D)_{ii}(\mathcal{Y}_D^\dagger \mathcal{Y}_D)_{jj}}\frac{(M_{i}^2-M_{j}^2) M_{i} \Gamma^0_{N_j}}{(M_{i}^2-M_{j}^2)^2+ (M_{i} \Gamma^T_{N_j})^2}   \,
\eta(T),
\end{eqnarray}
with
\begin{eqnarray}
\eta(T) &\equiv&x_p \left[1+f_B(E_H)-f_F(E_L)-2f_B(E_H)f_F(E_L)\right] .
\end{eqnarray}
Here, $f_B(E_H)$ and $f_F(E_L)$ denote the Bose--Einstein (BE) and
Fermi--Dirac (FD) distribution functions, respectively, which appear
from the finite--temperature cutting of the Higgs and lepton doublet
propagators. The quantity $\eta(T)$ encapsulates the thermal factor,
especially the cancellation arising from the compensation between
stimulated emission and Pauli blocking, as encoded in the identity
$f_B(E) - f_F(E) - 2 f_B(E) f_F(E) = 0.$ While an exact cancellation
needs the energies to be equal, in the rest frame of the decaying
particle, the energies of the internal states are given by
\begin{eqnarray}
E_H = \frac{M_i^2 + m_H^2(T) - m_L^2(T)}{2 M_i}, \quad
E_L = \frac{M_i^2 + m_L^2(T) - m_H^2(T)}{2 M_i}.
\end{eqnarray}
The exact cancellation is operative only when the thermal masses of the Higgs
and lepton doublets in the loop can be
neglected~\cite{Covi:1997dr}. 

\begin{figure}
    \centering
    \includegraphics[width=0.55\linewidth]{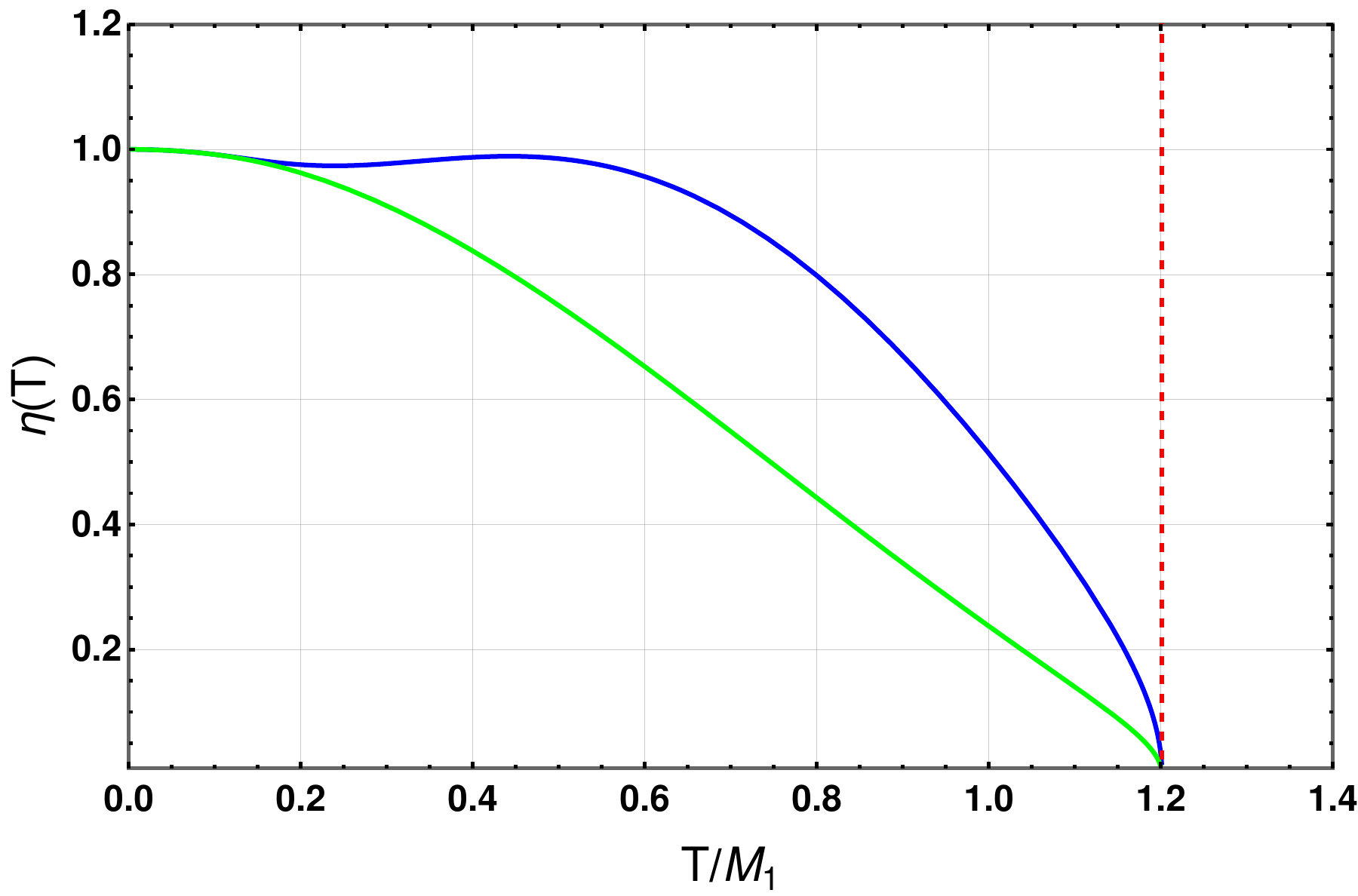}
\caption{Thermal correction factor to the CP-asymmetry, $\eta(T)$, as
  a function of temperature for $M_{1}=300\,\mathrm{GeV}$. The
  vertical red dashed line indicates the kinematic bound for the decay
  process $N_{1} \to H\,L_\alpha$. The blue curve represents
  $1+f_B(E_H) -f_F(E_L)$ rather than $1+f_B(E_H)
  -f_F(E_L)-2f_B(E_H)f_F(E_L)$~\cite{Garny:2009rv,Garny:2010nj,Seller:2024dkg}
  reflecting the modifications predicted by non-equilibrium field
  theory~\cite{Garny:2009rv,Garny:2009qn}.}
    \label{fig:etaT}
\end{figure}

Once thermal masses are consistently taken into account, the
cancellation is no longer exact, and the CP asymmetry receives thermal
corrections~\cite{Giudice:2003jh}.  In Fig.\,\ref{fig:etaT}, we
display the variation of $\eta(T)$ with $T/M_1$. At very low
temperatures, thermal contributions are insignificant on account of
the Boltzmann factor $\exp(-M_1/T)$. As the temperature approaches the
mass of the decaying particle, this suppression becomes less
pronounced, only to be finally replaced by another due to phase-space
effects. For example, the kinematic endpoint for the decay $N_1 \to H
L_\alpha$ is given by $T/M_1 = 1/(\sqrt{c_H}+\sqrt{c_L})\approx 1.2$
for our case, and is represented by the red dashed line.

Although the motion of the decaying particle through the plasma does,
in principle, introduce a velocity-dependent correction (to be
averaged over the thermal distribution), this
effect~\cite{Covi:1997dr, Giudice:2003jh} is small in the
  present context and can be neglected\footnote{Note that the
  anologue effect in $H \to N L$---accessible at high
  temperatures---could be non-negligible.}.

Greater rigour can be achieved through the use of non-equilibrium
quantum field theory (NQFT) based on the
Schwinger–Keldysh/Kadanoff–Baym formalism
\cite{Kadanoff:1962,Anisimov:2010dk,Garny:2009rv,Garny:2009qn,Seller:2024dkg,Dev:2017trv,Depta:2020zmy,Iso:2013lba,Dev:2017wwc}. This
top-down method yields results that differ from those obtained via
equilibrium thermal field theory, particularly in the statistical
factors that appear in the collision terms and the CP-violating
parameter. As shown in
refs.\cite{Garny:2009rv,Garny:2009qn,Kadanoff:1962}, the medium
corrections derived from NQFT are linear in the particle number
densities, in contrast to the quadratic dependence found in the
equilibrium thermal field theory treatments, and this different
density dependence in NQFT leads to an enhancement of the CP-violating
parameter~(see Appendix~\ref{app:nqft}). However, in the phenomenologically relevant regime where
the heavy Majorana neutrino mass $M_i$ is only of the order of the
temperature $T$, and considering massive leptons and Higgs doublets,
the phase space for decays is relatively small. Consequently, the
impact of these additional corrections on the final asymmetry is
subleading, especially at lower temperatures ($M_i < T$) where
sphaleron processes, which convert the lepton asymmetry into a baryon
asymmetry, remain active.

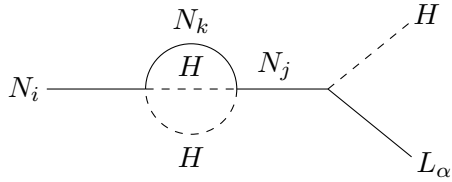
\begin{figure}[ht]
\centering
\begin{tikzpicture}[scale=1.0]

\node at (-0.8,0) {$N_i$};
\draw (-0.5,0) -- (0.8,0);

\draw (0.8,0) arc (180:0:0.6);            
\draw[dashed] (2.0,0) arc (0:-180:0.6);   

\draw[dashed] (0.8,0) -- (2.0,0);

\node at (1.4,0.9) {$N_k$};
\node at (1.4,-0.9) {$H$};
\node at (1.4, 0.3) {$H$};

\node at (2.5,0.3) {$N_j$};

\draw (2.0,0) -- (3.2,0);

\draw[dashed] (3.2,0) -- (4.3,0.9);   
\draw (3.2,0) -- (4.3,-0.9);       

\node at (4.5,1.0) {$H$};
\node at (4.6,-1.0) {$L_\alpha$};

\end{tikzpicture}

\caption{
Self-energy feynman diagram contributing to the CP asymmetry in the decay of heavy Majorana neutrinos, induced by the dimension-5 operator.} \label{fig:selfnnhh2}
\end{figure}

As for the contribution of the dimension-five operator $N^2H^2$
  to the CP asymmetry, not only is it suppressed by a factor of
  $\Lambda^{-2}$, but it also arises only at the two-loop level (see
Fig.\,\ref{fig:selfnnhh2}). Although its
propagator structure resembles that of a resonant self-energy
topology, any potential enhancement is offset by these suppressions,
rendering the contribution numerically subleading. It should
  also be noted that since self-energy
contributions are nonzero only in the presence of a mass splitting
between the heavy states (see Eq.~\eqref{eq:cp_asym}), the $N^2H^2$
operator yields a vanishing CP asymmetry when the coupling matrix
$c_{NH}$ is proportional to the identity.

In order to determine the resulting baryon asymmetry, one must solve
the coupled Boltzmann equations governing the evolution of the
heavy-neutrino abundances (in terms of time, or equivalently, the
ambient temperature $T$), as well as the total $B-L$ asymmetry.
Defining $z \equiv M_1/T$ and $Y_X \equiv n_X/s$ as the comoving
number density of species $X$ (with $s$ being the entropy density),
the evolution equations are given by:

\begin{eqnarray}\label{eq:boltz}
&& \frac{dY_{N_1}}{dz}=\frac{-z}{s H(M_1)}\left\{\left(\frac{Y_{N_1}}{Y_{N_1}^{eq}}-1\right)\Big(\gamma_{N_1}+2\gamma_{H,1}^s+4\gamma_{H,1}^t\Big)]\right\},   \nonumber \\
 && \frac{dY_{N_2}}{dz}=\frac{-z}{s H(M_1)}\left\{\left(\frac{Y_{N_2}}{Y_{N_2}^{eq}}-1\right)\Big(\gamma_{N_2}+2\gamma_{H,2}^s+4\gamma_{H,2}^t\Big)\right\},  \nonumber \\
    &&  \frac{dY_{B-L}}{dz}=\frac{-z}{s H(M_1)}\Biggr\{\left[\sum_{j=1}^{2} \frac{1}{2}\frac{Y_{B-L}}{Y_l^{eq}}+\epsilon_j\left(\frac{Y_{N_j}}{Y_{N_j}^{eq}}-1\right)\right]\gamma_{N_j}\nonumber \\
    &&\hspace*{5em} +\frac{Y_{B-L}}{Y_l^{eq}}\Big(2\gamma_{N,s}+2\gamma_{N,t}\Big)+\frac{Y_{B-L}}{Y_l^{eq}}\sum_{j=1}^{2}\left(2\gamma_{H,j}^{t}+\frac{Y_{N_j}}{Y_{N_j}^{eq}}\gamma_{H,j}^s\right)\Biggr\}.
  \label{eq:bolteq}
\end{eqnarray}
Here, $H(M_1)$ denotes the Hubble expansion rate evaluated at $T =
M_1$, and $s$ is the entropy density. The quantities
$\gamma_A$ represent the thermally
averaged decay/interaction rates. For example, $\gamma_{N_i}$ refers to
$N_i \leftrightarrow \ell H$, whereas the terms
$\gamma_{H,i}^{s}$ and $\gamma_{H,i}^{t}$ correspond to the
$\Delta L = 1$ Higgs-mediated scattering processes in the $s$- and
$t$-channels, respectively. Finally,
$\gamma_{N,s}$ and $\gamma_{N,t}$ refer to
the dominant $\Delta L = 2$ washout processes: $s$-channel scatterings
$l l \leftrightarrow H H$ and $t$-channel scatterings $l H
\leftrightarrow \bar{l} H^\dagger$, mediated by the heavy neutrinos. The sum over $j$ in the third equation incorporates the contributions from both heavy states. Explicit expressions for all the relevant reaction rates are analogous to the standard results of Refs.~\cite{Pilaftsis:1997jf,Plumacher:1996kc} and are collected in
Appendix~\ref{app:gamma_beexp}.


\subsection{Numerical analysis and results}
\label{sec:numerical}

As we have seen in the preceding section, the neutrino sector relevant
for leptogenesis corresponds to a complex symmetric $5\times5$ mass
matrix involving the three active neutrinos and the two heavy states
$N_{1,2}$, thereby rendering one light neutrino exactly massless. Once
oscillation data is reproduced, and the constraints from charged
lepton flavor violation imposed, the cosmological bounds are satisfied
automatically. The only remaining parameters relevant to leptogenesis
are the masses $M_{1,2}$ (which, of course are required to be equal)
and $(x_1, y_1$), {\em viz.}, the single complex angle of the
$2\times3$ Casas-Ibarra matrix $R$. To be specific, we choose two
benchmark points, namely
\[
M_1 = M_2 = 200\ {\rm GeV}  \qquad {\rm and} \qquad
M_1 = M_2 = 300\ {\rm GeV} \ .
\]

Since leptogenesis in our setup operates at temperatures $T \sim m_N$
and, thus, may lie above the electroweak scale, the requisite tiny
splitting of the heavy neutrino masses must already be realized in the
symmetric phase. The first contribution arises from
renormalisation group evolution and one-loop threshold corrections
within our low-energy EFT with cutoff $\Lambda \sim
1\,\text{TeV}$. Even if the tree-level Majorana mass matrix $m_R$ is
exactly degenerate at the cutoff scale, radiative corrections from
Yukawa interactions generate off-diagonal entries. Solving the RG
equation for the heavy-neutrino mass matrix, as in Eq.~(2.12), and
running from $\Lambda$ down to $m_N$ yields a zero-temperature
one-loop splitting parametrically given by~\cite{Pilaftsis:2005rv}

\begin{eqnarray}
\Delta M_{12}^{\text{(1-loop)}} \sim \frac{m_N}{8\pi^2} \left| \Re\left[ (\mathcal{Y}_D^\dagger \mathcal{Y}_D)_{12} \right] \right| \ln\left( \frac{\Lambda}{m_N} \right).    
\end{eqnarray}
For $\Lambda \sim 1\,\text{TeV}$ and $m_N$ near the electroweak scale, the logarithmic factor is of order unity, making this contribution
marginally smaller than the decay widths $\Gamma_{N_i}$ (see Eq.\,\eqref{N_width}).

In addition to this zero-temperature effect, finite-temperature
self-energy corrections arising from Yukawa interactions induce
temperature-dependent contributions to the heavy-neutrino mass
matrix. In the regime $M_i \simeq M_j \gtrsim T$, these effects lift
the degeneracy and generate an effective mass splitting even within an
otherwise degenerate pair, to the tune of~\cite{Pilaftsis:2003gt}

\begin{equation}
\Delta M_{12}(T) \sim \frac{T^2}{16 \, m_N} \, \Re\left[(\mathcal{Y}_D^\dagger \mathcal{Y}_D)_{12}\right] \ .
\end{equation}
For temperatures relevant to leptogenesis, this correction is of the same order as the one-loop RG-induced splitting.
Both the splittings are included in our analysis.  and, together, they respect $\Delta M_{12} \sim \Gamma_N$. Resonance can, therefore, be dynamically realized in the symmetric phase through a combination of RG-induced and thermal effects.

After electroweak symmetry breaking, the Higgs VEV induces
active–sterile mixing through the off-diagonal Dirac entries of the
full seesaw mass matrix. Diagonalization of the complete $(\nu_L,
N^c)$ system then generates an additional vacuum contribution to the
heavy-neutrino spectrum. As the temperature drops, the thermal
correction is suppressed, and the residual splitting is governed
predominantly by this mixing. The resonant condition can thus be
satisfied prior to symmetry breaking, while the late-time spectrum is
fixed by seesaw diagonalization.

In the presence of a strong first-order electroweak phase transition
with a relatively low nucleation temperature $T_N$, the Universe
remains in the symmetric phase down to $T_N$, permitting
heavy-neutrino decays to generate a $B-L$ asymmetry while sphalerons
remain active~\cite{Phong:2020ybr}.  Because the phase transition
delays sphaleron decoupling relative to the SM case, baryon-number conversion can remain efficient even
below the usual SM sphaleron freeze-out temperature. Once the broken
phase bubbles nucleate and expand, sphaleron processes become
exponentially suppressed inside the bubbles, thereby preserving the
converted baryon asymmetry.

\begin{figure}
    \centering
    \includegraphics[width=0.49\linewidth]{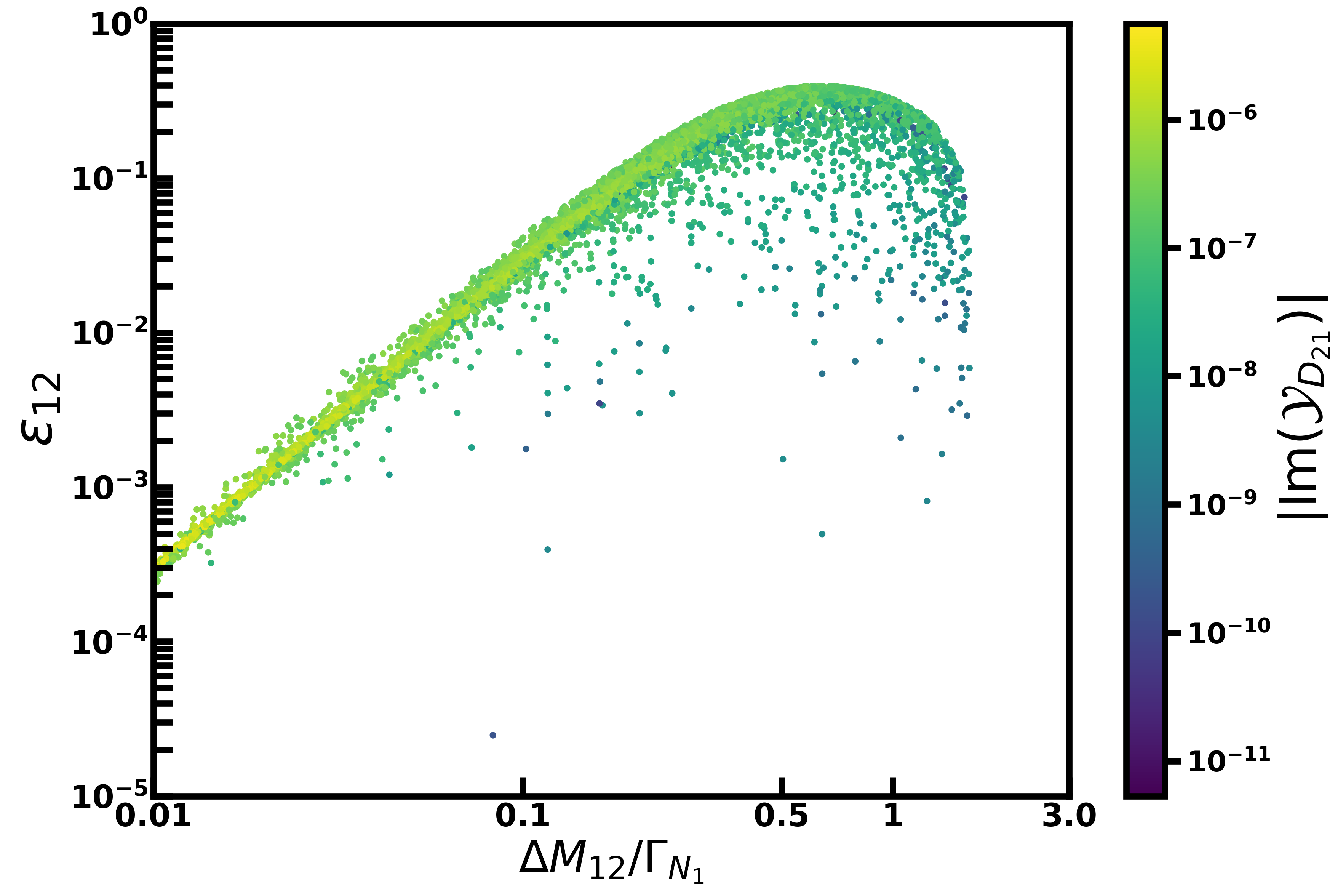}
    \includegraphics[width=0.49\linewidth]{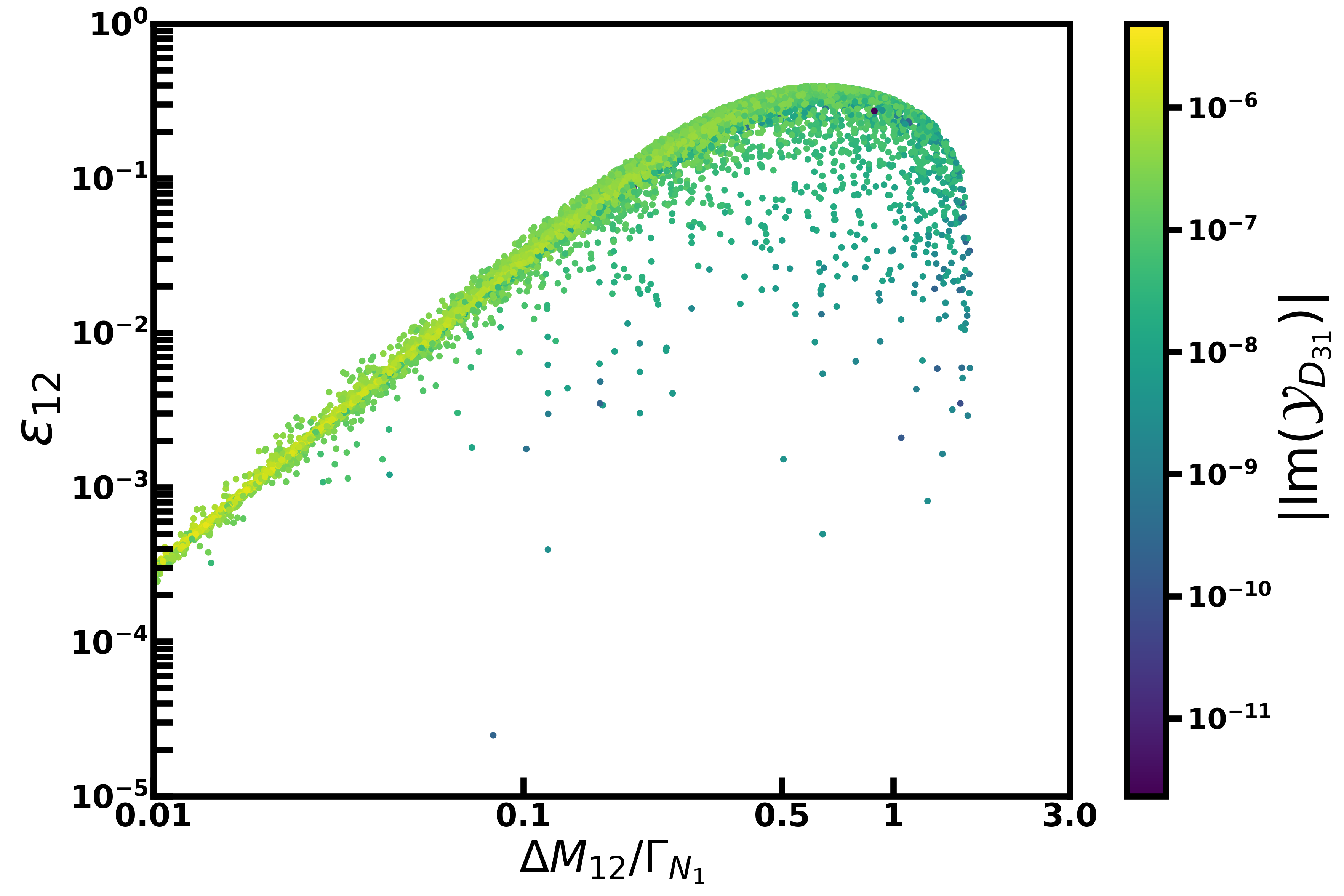}
\caption{Scatter plots of the CP asymmetry parameter $\epsilon_{12}$
  as a function of the ratio of heavy-neutrino mass splitting and
  decay width, $\frac{\Delta {M_{12}}}{\Gamma_{N_1}}$.  Each point
  corresponds to a parameter configuration consistent with neutrino
  oscillation and charged lepton flavor violation constraints.  The
  color bar indicates the magnitude of the imaginary part of Yukawa
  couplings. Here, $(M_1,M_2)\approx (300,300)$ GeV, and temperature
  $T=200$ GeV.}
    \label{fig:eps_vs_split}
\end{figure}
The resonant behaviour of the CP asymmetry parameter $\epsilon_{12}$,
as a function of the ratio $\Delta M_{12}/\Gamma_{N_1}$ is illustrated
in Fig.\,\ref{fig:eps_vs_split}. The splitting shown here includes
temperature-dependent contributions induced by Yukawa interactions at
$T=200$~GeV. Each point corresponds to a viable scan configuration
satisfying neutrino oscillation and charged lepton flavor violation
constraints. The color coding indicates the magnitude of the imaginary
part of the relevant Yukawa coupling.

A clear resonant structure is observed. The CP asymmetry increases as
the splitting approaches the regime $\Delta M_{12} \sim \Gamma_{N_1}$
and reaches its maximal values in this region (more specifically,
around $\Delta M_{12} \sim \Gamma_{N_1}/2$). For larger values of this
ratio, the asymmetry decreases smoothly, reflecting the suppression of
the self-energy enhancement as the spectrum moves away from
quasi-degeneracy.

With the mass spectrum fixed, we solve the coupled Boltzmann equations
for the heavy-neutrino abundances and the total $B-L$ asymmetry,
including decays, inverse decays, and the relevant $\Delta L=1$
scattering and washout processes discussed earlier. The final baryon
asymmetry is obtained through sphaleron conversion, and
\begin{equation}
Y_B=\left(\frac{8n_f+4n_h}{22n_f+13n_h}\right)Y_{B-L}(z_{\rm sph}),
\end{equation}
where $n_f$ and $n_h$ denote the number of fermion generations and
Higgs doublets, respectively and $z_{\rm sph}=M_1/T_{\rm sph}$ and
$T_{\rm sph}\simeq150~\mathrm{GeV}$~\cite{Burnier:2005hp}.

\begin{table}[h!]
\centering
\small
\begin{tabular}{|c|c|c|c|}
\hline
Scenario & $(M_1,M_2)$ [GeV] & $\Delta M_{12}$ [GeV] & Yukawa order ($\mathcal{Y}_D$) \\
\hline
A & $(200,\,200)$ & $10^{-10}\text{--}10^{-11}$ & $\mathcal{O}(10^{-6})$ \\
\hline
B & $(300,\,300)$ & $10^{-10}\text{--}10^{-11}$ & $\mathcal{O}(10^{-6})$ \\
\hline
\end{tabular}
\caption{Representative benchmark configurations in the two–right-handed-neutrino framework. }
\label{tab:benchmarks}
\end{table}

\begin{figure}
    \centering
    \includegraphics[width=0.49\linewidth]{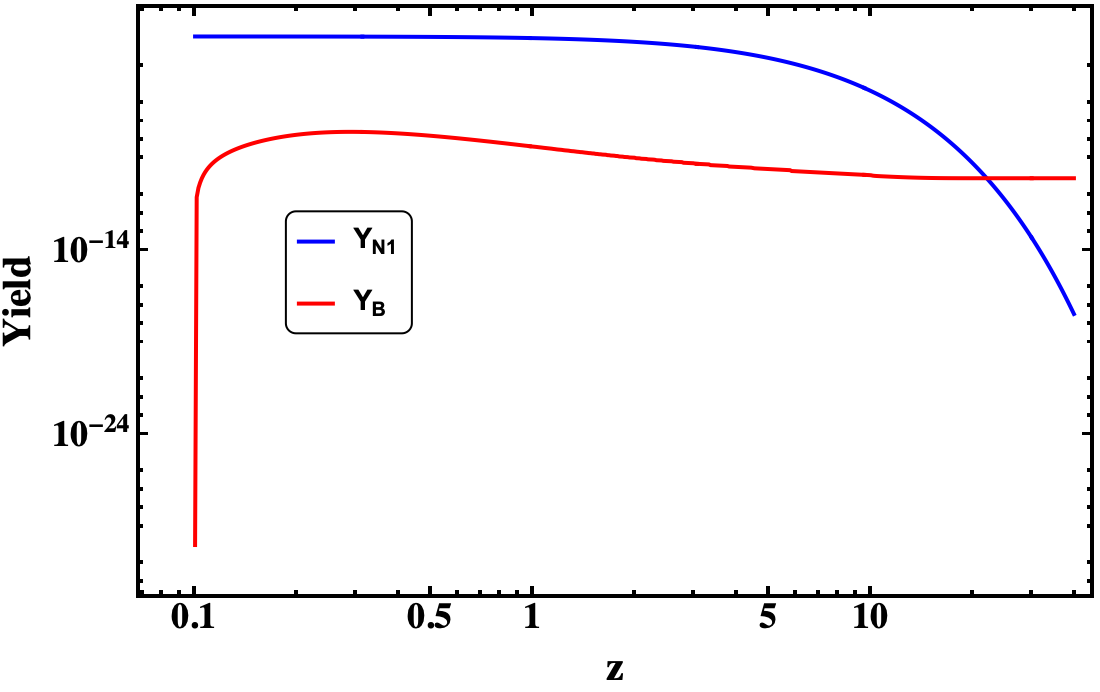}
    \caption{ Evolution of $Y_{N_1}$ and $Y_B$ as a
      function of $z = M_1/T$ With thermal and 1-loop RGE induced mass splittings.  } \label{fig:yb_eol}
\end{figure}

The evolution of the heavy-neutrino abundances and the resulting
baryon asymmetry is shown in Fig.\,\ref{fig:yb_eol}. At small values of
\(z\), corresponding to the high-temperature regime, the heavy
neutrinos remain close to thermal equilibrium and the generated
asymmetry is strongly suppressed. As the temperature decreases, the
system gradually departs from equilibrium. In this regime,
finite-temperature effects together with the one-loop contribution
generate a nonzero mass splitting between the heavy neutrino states,
leading to the onset of CP asymmetry generation already in the
symmetric phase. Around the electroweak transition, the splitting is
further affected by the Higgs-induced Yukawa contribution, and the
resonance condition is efficiently realized, resulting in a rapid
growth of \(Y_B\) around \(z \sim \mathcal{O}(1)\).  At later times,
as the temperature drops further, the relevant washout processes
become inefficient and the baryon asymmetry freezes to a constant
value. The final asymmetry is consistent with the observed baryon
asymmetry of the Universe.

\begin{figure}
    \centering
    \includegraphics[width=0.5\linewidth]{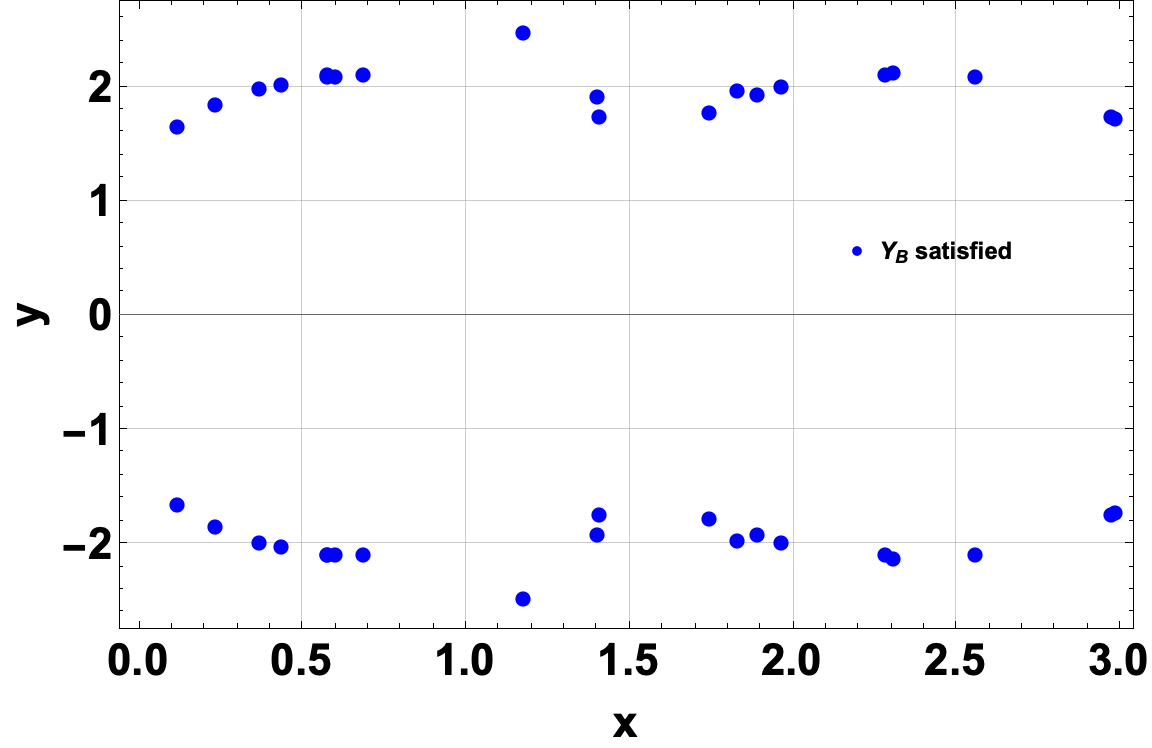}
    \caption{
Parameter space in Casus Ibarra parameter $(x,y)$ plane that satisfy neutrino mass, lepton flavor violation, and observed baryon asymmetry as well.
} \label{fig:Ybscanplot}
\end{figure}

Fig.\,\ref{fig:Ybscanplot} shows the result of the numerical scan
over the complex Casas--Ibarra parameter $ z=x+iy $ for the benchmark
choice $ M_1=M_2=300~\mathrm{GeV}.  $ The scan is performed over the
indicated range in the $(x,y)$ plane, while imposing the neutrino
oscillation and charged lepton flavor violation constraints. The red
points denote the parameter choices that, in addition, reproduce the
observed baryon asymmetry of the Universe. The figure therefore
illustrates the regions in the scanned parameter space where all these
requirements are simultaneously satisfied. Representative benchmark
points are summarized in Tab.\,\ref{tab:benchmarks}.

In more general realizations featuring an additional heavy neutrino
that also mixes with the active sector, operator-induced self-energy
effects can arise. Prior to electroweak symmetry breaking these enter
only at two-loop order, whereas after symmetry breaking they may
induce one-loop self-energy corrections through operator-generated
mass insertions. The quantitative impact is spectrum dependent, since
the relevant loop functions depend on the mass of the additional
state: if it is sufficiently light, kinematic suppression can be
milder and the operator-induced contribution correspondingly less
suppressed. In special flavour textures and in the absence of strong
phase-space suppression, such effects can become comparable to the
Yukawa-induced contributions. However, this requires a non-minimal
spectrum and typically involves a degree of tuning beyond the
two–right-handed-neutrino framework considered here.

\section{Dark matter phenomenology}
\label{sec:dm}
In the present framework, one of the neutral fermions naturally
emerges as a viable DM candidate. As mentioned earlier, the
third-generation heavy neutrino, $N_3$, is odd under a discrete $Z_2$
symmetry, while all other fields in the theory are taken to be $Z_2$
even. This discrete symmetry forbids all decay channels of $N_3$ into
the SM states, thereby ensuring its absolute stability on cosmological
timescales.

The relevant part of the Lagrangian, upto mass dimension six,
governing the dark matter sector is given by
\begin{equation}
\mathcal{L}_{\rm DM}
=
\overline{N_3}\, i\slashed{\partial}\, N_3
-
\left(
\frac{m_{R_3}}{2}\, \overline{N_{3}^{\,c}}\, N_3 
+
\frac{c_{NH_{33}}}{2\Lambda}\, \overline{N_3^{\,c}} N_3 \, H^\dagger H 
+ \text{h.c.}
\right)
+
\frac{1}{\Lambda^2}
\sum_i C_i\, \mathcal{O}_i \, ,
\label{eq:LDM}
\end{equation}
where, for the sake of simplicity, the WCs $c_{NH_{33}}$ and $C_i$ (corresponding to the dimension-5 and dimension-6 operators respctively) are taken to be real.

At dimension six, several operators can contribute non-trivially to
dark matter annihilation, elastic scattering, and indirect
signatures. The operators relevant for our analysis
are~\cite{Liao:2016qyd}
\begin{eqnarray}
&& \mathcal{O}_{HN_3}
=
(\overline{N_3}\gamma^\mu N_3)\,
(H^\dagger i\overleftrightarrow{D}_\mu H), \qquad
\mathcal{O}_{LN_3}
=
(\overline{N_3}\gamma^\mu N_3)\,
(\overline{L}\gamma_\mu L), \nonumber\\
&& \mathcal{O}_{QN_3}
=
(\overline{N_3}\gamma^\mu N_3)\,
(\overline{Q}\gamma_\mu Q), \hspace{14.5mm}
\mathcal{O}_{eN_3}
=
(\overline{N_3}\gamma^\mu N_3)\,
(\overline{e}\gamma_\mu e), \nonumber\\
&& \mathcal{O}_{uN_3}
=
(\overline{N_3}\gamma^\mu N_3)\,
(\overline{u}\gamma_\mu u), \hspace{17mm}
\mathcal{O}_{dN_3}
=
(\overline{N_3}\gamma^\mu N_3)\,
(\overline{d}\gamma_\mu d), \nonumber\\
&& \mathcal{O}_{N_3N_3}
=
(\overline{N_3}\gamma^\mu N_3)\,
(\overline{N_i}\gamma_\mu N_j) + \text{h.c.}(i\neq j, j>i ),
\quad i,j=1,2 .
\end{eqnarray}
We adopt a minimal effective description and retain only operators
involving the dark fermion $N_3$ and the Higgs doublet. Operators
coupling $N_3$ directly to the SM fermions are set to zero, since they
are not essential for the dark matter phenomenology discussed here and
are subject to stronger flavor and collider constraints.

\begin{figure}
    \centering
    \includegraphics[width=0.49\linewidth]{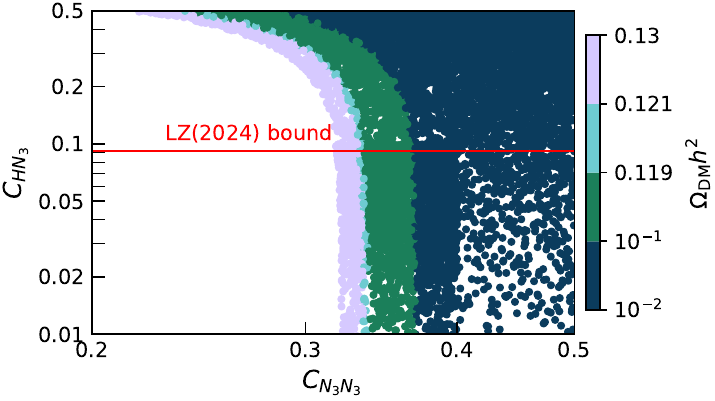}
    \includegraphics[width=0.49\linewidth]{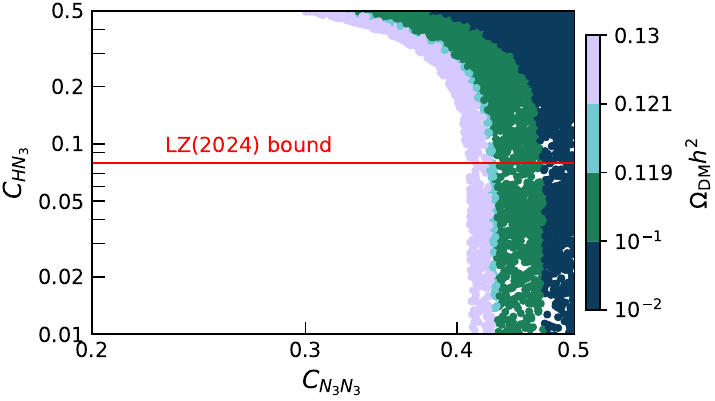}
\caption{Dark matter relic abundance as a function of the effective
  Wilson coefficient controlling the $N_3$ interactions. The left
  panel corresponds to $(m_{N_1}, m_{N_2},
  m_{N_3})=(300,300,400)\,\mathrm{GeV}$, while the right panel shows
  the case $(200,200,300)\,\mathrm{GeV}$. In both panels the cutoff
  scale is fixed to $\Lambda=1~\mathrm{TeV}$, with
  $c_{NH_{33}}=10^{-3}$ and $C_{N_{322}}=C_{N_{312}}=C_{N_{311}}\equiv
  C_{N_3N_3}$. The horizontal band indicates the observed dark matter
  relic density.}
    \label{fig:relic}
\end{figure}

\begin{figure}
    \centering
    \includegraphics[width=0.7\linewidth]{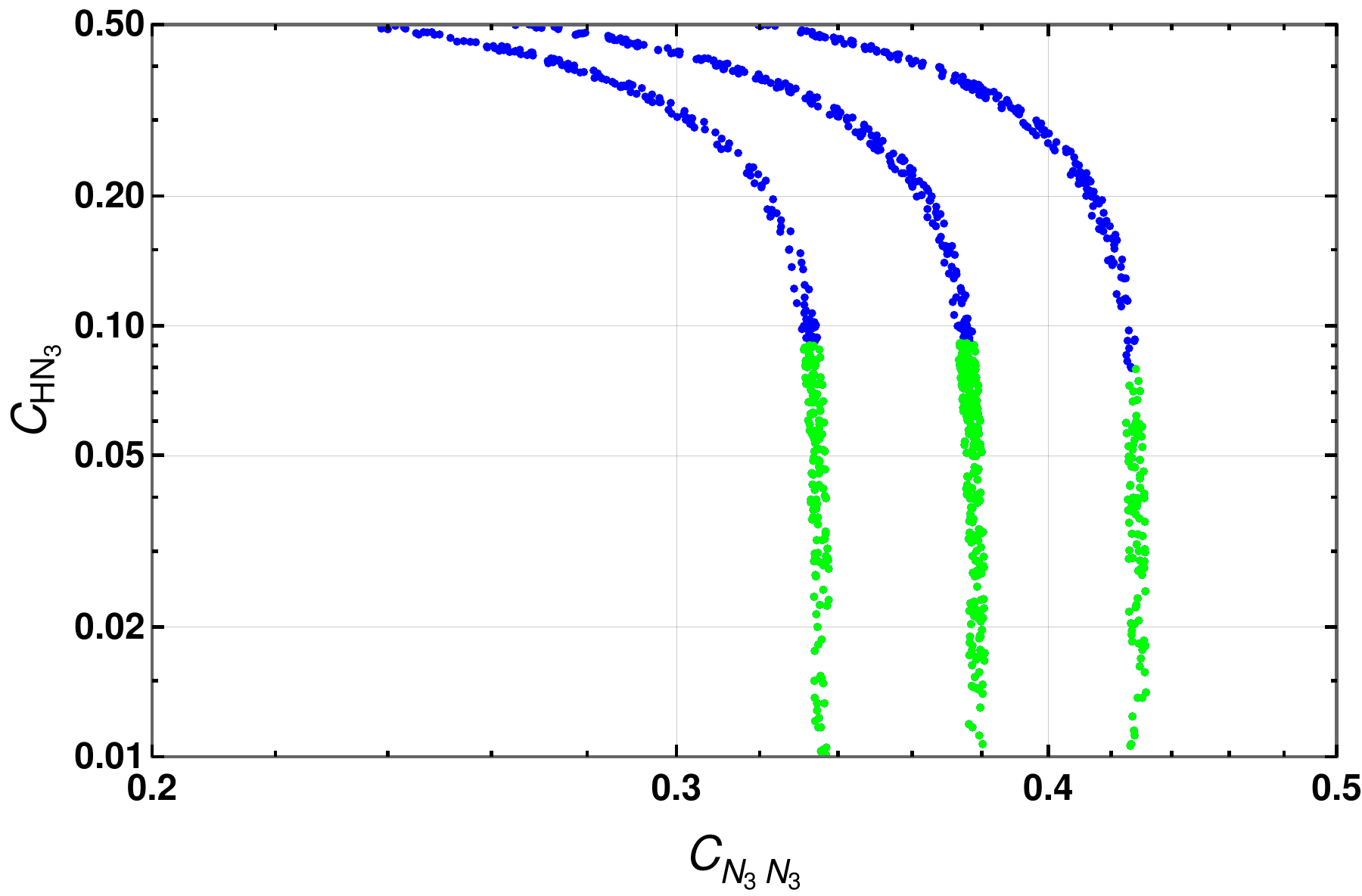}
\caption{ Allowed regions in the $(C_{HN_3},\, C_{N_3N_3})$ parameter
  plane that reproduce the observed dark matter relic abundance,
  $0.119 \le \Omega_{\rm DM} h^2 \le 0.121$. The three curves
  correspond to the heavy-neutrino mass configurations $(m_{N_1},
  m_{N_2}, m_{N_3}) = (300,300,400)\,\mathrm{GeV}$ (left),
  $(200,200,400)\,\mathrm{GeV}$ (middle), and
  $(200,200,300)\,\mathrm{GeV}$ (right). In all cases the cutoff scale
  is fixed to $\Lambda = 1~\mathrm{TeV}$, with $c_{NH_{33}} = 10^{-3}$
  and $C_{N_{322}} = C_{N_{312}} = C_{N_{311}}\equiv C_{N_3N_3}$. The
  blue points are excluded by current $2\sigma$ limits from
  spin-dependent dark matter--nucleon scattering experiments.  }
    \label{fig:placeholder}
\end{figure}

The imposed $Z_2$ symmetry forbids the renormalisable Yukawa
interaction $\overline{L}\tilde{H} N_3$ as well as mixed effective
operators involving $N_{1,2}$, ensuring that $N_3$ does not decay or
coannihilate with lighter sterile states. Consequently, the relic
abundance of $N_3$ is set entirely by thermal freeze-out. The relevant annihilation and scattering
processes are evaluated using
\texttt{micrOMEGAs}~\cite{Belanger:2004yn,Alguero:2023zol}.

We first consider the scenario in which only the dimension-5 operator
proportional to $c_{NH_{33}}$ is present.  This operator induces
Higgs-mediated annihilation as well as spin-independent elastic
scattering off nuclei.  We find that satisfying the stringent limits
from the LZ-2024 experiment on spin-independent
scattering~\cite{LZ:2024zvo} requires $c_{NH_{33}}$ to be strongly
suppressed, rendering this operator alone insufficient to reproduce
the observed dark matter relic abundance.

Considering the direct detection bound, we fix $c_{NH_{33}}=0.001$ in the following and extend the analysis by
including dimension-6 operators.
The relic density obtained in this regime is illustrated in
Fig.\,\ref{fig:relic} for two representative mass choices,
$(m_{N_1},m_{N_2},m_{N_3})=(300,300,400)\,\mathrm{GeV}$ and
$(200,200,300)\,\mathrm{GeV}$.
In both cases, the relic abundance is controlled almost entirely by dimension-6
interactions, while the contribution from the suppressed dimension-5 operator is
negligible.

Among the dimension-6 operators, two operators are phenomenologically important:
the Higgs-current operator $\mathcal{O}_{HN_3}$ and the neutrino four-fermion operator
$\mathcal{O}_{N_3N_3}$.
The role of $\mathcal{O}_{HN_3}$ is primarily to induce electroweak-current interactions and,
importantly, spin-dependent elastic scattering off nuclei, so that LZ-2024~\cite{LZ:2024zvo} bounds
directly enforce an upper limit on $C_{HN_3}$.

The dominant contribution to the relic abundance instead originates from the
four-fermion operator $\mathcal{O}_{N_3N_3}$.
This operator mediates dark matter annihilation into lighter neutral fermions,
\begin{equation}
N_3 N_3 \;\to\; N_i N_j \,, \qquad i,j=1,2 \, ,
\end{equation}
which constitutes the leading annihilation channel during thermal
freeze-out.  In our numerical analysis, the masses of $N_1$ and $N_2$
are taken to be degenerate. Under this assumption, it is natural to
treat the corresponding couplings as equal, and we therefore set
$C_{N_{322}} = C_{N_{312}} = C_{N_{311}} \equiv C_{N_3N_3}$.  As a
result, the mixed final state $N_3 N_3 \to N_1 N_2$ receives
contributions from two distinct contractions and therefore contributes
approximately twice as much as the diagonal channels $N_3 N_3 \to N_1
N_1$ and $N_3 N_3 \to N_2 N_2$.  The total annihilation cross section
is thus dominated by these neutrino final states.
The resulting parameter space consistent with the Planck $2\sigma$ relic density
interval,
$0.119 \le \Omega_{\rm DM} h^2 \le 0.121$,
is shown in Fig.\,\ref{fig:placeholder} for three heavy-neutrino mass hierarchies,
$(300,300,400)\,\mathrm{GeV}$, $(200,200,400)\,\mathrm{GeV}$, and
$(200,200,300)\,\mathrm{GeV}$.

For each mass configuration, the relic density constraint selects a
narrow band in the $(C_{HN_3},C_{N_3 N_3})$ plane.  The coefficient
$C_{HN_3}$ is primarily constrained by spin-dependent direct detection
limits, shown by the blue excluded regions, and has only a subleading
impact on the relic abundance.  In contrast, the four-fermion
coefficient $C_{N_3 N_3}$ is essentially unconstrained by direct
detection experiments and is fixed predominantly by the freeze-out
condition to be of order $\mathcal{O}(0.1)$.

Comparing the different mass hierarchies, heavier dark matter masses shift the
allowed region toward larger values of $C_{HN_3}$, as a stronger effective coupling
is required to compensate for phase-space suppression in annihilation.
The surviving green regions therefore identify the phenomenologically viable
parameter space in which the observed relic density, direct detection bounds, and
effective field theory validity are simultaneously satisfied.
In this minimal scenario, the relic abundance is set mainly by annihilation into
neutrino final states, while annihilation into Standard Model particles, such as
$t\bar{t}$, provides a subdominant contribution driven by
$\mathcal{O}_{HN_3}$.

\section{Summary and conclusion}
\label{sec:concl}

The observed baryon asymmetry of the Universe cannot be explained within the
Standard Model. This motivates new sources of CP violation together with new
out-of-equilibrium dynamics. In this work, we studied a neutrino-extended
effective field theory of the Standard Model, including higher-dimensional
operators up to dimension six, and showed that it can simultaneously accommodate
a strong first-order electroweak phase transition, low-scale resonant
leptogenesis, and a viable fermionic dark matter candidate.

We first analyzed the electroweak phase transition in the presence of the
dimension-six operators. Among them, the pure Higgs operator \(O_H\) provides
the dominant effect in strengthening the transition, while the contributions
from \(C_{\rm kin}\) and \(C_{tH}\) are subleading. For representative
parameter choices compatible with electroweak precision constraints, we found
that the transition strength \(v_c/T_c\) can lie in the range
\(0.8\!-\!2.9\), with a substantial region satisfying the usual strong
first-order criterion \(v_c/T_c \gtrsim 1\). The dimension-five
neutrino--Higgs operator also contributes to the finite-temperature potential
through the field-dependent heavy-neutrino masses. Although this effect is
subdominant compared with the Higgs-sector operator, it provides a direct
connection between the neutrino sector and the phase-transition dynamics. For
the benchmark heavy-neutrino spectra considered in our analysis, this operator
induces only mild quantitative shifts and does not alter the conclusion that a
strong first-order transition can be realized.

We then computed the stochastic gravitational-wave signal associated with the
phase transition. The spectrum is dominated by the sound-wave contribution,
while magnetohydrodynamic turbulence gives a subleading high-frequency tail and
the bubble-collision contribution is negligible for non-runaway walls. In the
benchmark scenarios studied here, the peak frequency lies in the mHz to
sub-Hz range, placing the signal within the projected sensitivity of future
space-based interferometers such as LISA, BBO, and DECIGO, depending on the
details of the phase-transition parameters. The gravitational-wave prediction is
therefore controlled mainly by the Higgs-sector operators that determine the
transition strength, with the neutrino-sector operator affecting the signal only
indirectly through its mild modification of the effective potential.

The leptogenesis sector exhibits a complementary structure. Since the heavy
Majorana neutrinos lie near the electroweak scale, leptogenesis occurs at
temperatures \(T \sim m_N\), and the required tiny mass splitting of the
quasi-degenerate pair must already be generated in the symmetric phase. In our
framework, this splitting receives two relevant contributions. The first is the
zero-temperature one-loop contribution generated by renormalization-group
running and threshold effects in the low-energy EFT, even when the tree-level
Majorana mass matrix is exactly degenerate at the cutoff scale. The second is a
finite-temperature self-energy correction induced by Yukawa interactions, which
produces a thermal splitting of the form
\[
\Delta M_{12}(T)\sim \frac{T^2}{16m_N}\,
\mathrm{Re}\!\left[(Y_D^\dagger Y_D)_{12}\right].
\]
At temperatures relevant for leptogenesis, this thermal
contribution is of the same Yukawa order as the one-loop splitting, and must
  be included in the analysis. After electroweak
symmetry breaking, the Higgs vacuum expectation value induces
active--sterile mixing through the full seesaw mass matrix, adding a
further vacuum contribution to the heavy-neutrino spectrum.  Taken
together, these effects dynamically realize the resonance condition
\(\Delta M_{12}\sim \Gamma_N\).

We also explore the effects of the thermal masses of the Higgs and
lepton doublets and of the motion of the decaying heavy neutrino
through the plasma. Both corrections are numerically tiny in the
parameter region of interest and do not qualitatively modify our
results.

Solving the Boltzmann equations with decay, inverse-decay, and the
dominant washout processes, we found that the observed baryon
asymmetry can be reproduced for quasi-degenerate heavy-neutrino masses
at the few-hundred-GeV scale while remaining consistent with neutrino
oscillation data and charged-lepton-flavor violation constraints. In
the Casas--Ibarra scan, only restricted regions of the \((x,y)\) plane
simultaneously satisfy all requirements, illustrating the selective
nature of the resonant condition. In the successful region, the
required Yukawa couplings are typically in the range
\(\mathcal{O}(10^{-7}\!-\!10^{-6})\), while the CP asymmetry is
resonantly enhanced once the dynamically generated splitting
approaches the decay width.  The resulting baryon asymmetry freezes to
a constant value after the washout processes become inefficient. In
the presence of a strong first-order electroweak phase transition, the
late nucleation of broken-phase bubbles further helps preserve the
converted baryon asymmetry by suppressing sphaleron processes inside
the bubbles.

The same EFT setup also admits a minimal fermionic dark matter
sector. By stabilizing the third singlet fermion with a discrete
symmetry, its Yukawa mixing with the active neutrinos is forbidden and
cosmological stability is ensured. The relic abundance is obtained
through thermal freeze-out driven by higher-dimensional
interactions. While the dimension-five
Higgs-portal interaction is strongly constrained by direct-detection
limits, viable parameter space opens once the
dimension-six Higgs-current operator is included, allowing the
observed dark matter abundance to be reproduced without violating
present bounds.

In summary, this work presents a unified framework in which the Higgs-sector
operators dominantly govern the electroweak phase transition and the associated
gravitational-wave signal, while the neutrino-sector operator provides the
structure required for low-scale resonant leptogenesis and feeds only mildly
into the thermal potential. A central result is that the tiny mass splitting
required for resonant enhancement is generated dynamically in the symmetric
phase through the combined effect of the zero-temperature one-loop
RG/thresh\-old contribution and the finite-temperature Yukawa-induced
self-energy correction, with thermal mass effects included consistently in the
decay kinematics and CP asymmetry. The framework therefore correlates
electroweak phase-transition physics, gravitational waves, neutrino mass
generation, charged-lepton-flavor violation, baryogenesis, and fermionic dark
matter within a single neutrino-extended EFT description. This makes it a
promising setup for future tests in gravitational-wave observatories, flavor
experiments, and dark-matter searches.

\section*{Acknowledgment}
DC acknowledges the ANRF, Government of India, for support through the
project CRG/\allowbreak2023/008234 and the IoE, University of Delhi
grant IoE/2025-26/12/FRP. JD acknowledges the ANRF (formerly Science and
Engineering Research Board (SERB)), Government of
India, for the national postdoctoral fellowship (NPDF) grant PDF/2023/0015. TS is supported by the DST, Government of India through the DST INSPIRE Faculty Fellowship (DST/INSPIRE/04/2024/004616). The work is also supported by the National Natural Science Foundation of China under Grant Nos. 12475094, 12135006, and 12075097.

\appendix
\section{Coefficients of \ensuremath{V_{\rm ct}(h)}}\label{app:ct}
The coefficients of $V_{\rm ct}(h)$ given in Eq.~\eqref{eq:ctpot} can be calculated using the following renormalization conditions:
\begin{eqnarray}
 \left\{\frac{d}{dh},\frac{d^2}{dh^2}\right\} \Big(V^{\rm CW}_{\rm 1-loop}(h,T=0)+V_{\rm ct}(h)\Big) \Big|_{h=v}=0.
 \end{eqnarray}
 Using the above conditions, the coefficients can be obtained as
\begin{eqnarray}
    &&\delta\mu_H^2 = -\frac{1}{2v}\left(3 \frac{dV^{\rm CW}}{d h} -v \frac{d^2V^{\rm CW}}{d h^2} \right)\Big|_{h=v},\\
    &&\delta\lambda_H = -\frac{1}{2v^3}\left(v \frac{d^2V^{\rm CW}}{d h^2} -\frac{dV^{\rm CW}}{d h} \right)\Big|_{h=v}.
\end{eqnarray}
Here, we have used the short-hand notation $V^{\rm CW}\equiv V^{\rm CW}_{\rm 1-loop}(h,T=0)$. 

\subsection*{CP asymmetry in NQFT framework}
\label{app:nqft}
The resonant contribution to the CP asymmetry in the NQFT framework can be written as \cite{Frossard:2012pc}: 
\begin{eqnarray}\label{eq:cp_asymNE}
\epsilon_{ij}(T)&=& \dis
\frac{\text{Im}[\mathcal{Y}_D^\dagger \mathcal{Y}_D]_{ij}^2}{(\mathcal{Y}_D^\dagger \mathcal{Y}_D)_{ii}(\mathcal{Y}_D^\dagger \mathcal{Y}_D)_{jj}}\frac{(M_{i}^2-M_{j}^2) M_{i} \Gamma^0_{N_j}}{(M_{i}^2-M_{j}^2)^2+ \left(\frac{\Gamma^0_{N_j}}{M_{i}} L.q_{N_i} \right)^2} \left(\frac{  p_L.L}{p_L.q_{N_i}}\right) ,
\end{eqnarray}
with the Lorentz four vector
\begin{eqnarray}
L^\mu &=& 16\pi\int d\Pi_H(k_H) d\Pi_L(p_L) (2\pi)^4 \delta^4(q_{N_i} - k_H - p_L)\, p_L^\mu  \big[1 + f_B(E_H) - f_F(E_L)\big],\nonumber
\end{eqnarray}
where \( d\Pi_a(p) = \frac{d^3 p}{(2\pi)^3\, 2E_p} \) is the invariant phase space measure. In the rest frame of the decaying particle, scalar dot products are simplified as
\begin{eqnarray}
    \frac{p_L.L}{q_{N_i}.p_L} =\frac{q_{N_i}.L}{M_{i}^2} = x_p  \big[1 + f_B(E_H) - f_F(E_L)\big].
\end{eqnarray}
Using the above two equations, one can get 
\begin{eqnarray}
\epsilon_{ij}(T)&=& \dis
\frac{\text{Im}[\mathcal{Y}_D^\dagger \mathcal{Y}_D]_{ij}^2}{(\mathcal{Y}_D^\dagger \mathcal{Y}_D)_{ii}(\mathcal{Y}_D^\dagger \mathcal{Y}_D)_{jj}}\frac{(M_{i}^2-M_{j}^2) M_{i} \Gamma^T_{N_j}}{(M_{i}^2-M_{j}^2)^2+ \left(m_{N_i}\Gamma^T_{N_j}\right)^2},
\end{eqnarray}
with the effective decay width of the heavy neutrino $N_j$
\begin{eqnarray}
   \Gamma^T_{N_j} = \Gamma^0_{N_j} x_p  \left[1 + f_B(E_H) - f_F(E_L)\right],
\end{eqnarray}
where $\Gamma^0_{N_j}= \frac{(\mathcal{Y}_D^\dagger \mathcal{Y}_D)_{ii} M_{j}}{8\pi} $ is its vacuum decay width.
\section{Contribution to CLFV processes}\label{app:lfvexp}
\subsection*{Lepton radiative decay}
The branching ratio for \(\ell^\prime \to \ell\gamma\) is given by~\cite{Dinh:2013tvc,Casas:2001sr}
\begin{equation}
B(\ell^\prime\to \ell\gamma)=\frac{3\alpha_{\text{em}}}{32\pi}\,|T|^2,
\label{eq:mueg}
\end{equation}
where \(\alpha_{\text{em}}\) is the fine‑structure constant and
\begin{align}
T&=\sum_{j=1}^{3}\big[(1+\eta)U\big]_{\ell^\prime j}^{*}\,
\big[(1+\eta)U\big]_{\ell j}\,
G\!\left(\frac{m_{\nu_j}^2}{m_W^2}\right) \nonumber\\
&\quad+\sum_{k=1}^{2}\big(RV\big)_{\ell^\prime k}^{*}\,
\big(RV\big)_{\ell k}\,
G\!\left(\frac{M_k^2}{m_W^2}\right),
\label{eq:T}
\end{align}
with the loop function
\begin{equation}
G(x)=\frac{10-43x+78x^2-49x^3+4x^4+18x^3\ln x}{3(x-1)^4}.
\label{eq:G}
\end{equation}
The first sum runs over the three light neutrinos (masses \(m_{\nu_j}\)), the second over the two heavy Majorana neutrinos (masses \(M_k\)). The matrices \(U,\eta\) encode the PMNS mixing, non‑unitarity corrections. Here, $RV = -i U D_{m_\nu}^{1/2} R\, D_{{m_N}}^{-1/2}$ represents the active heavy neutrino mixing. Here, the diagonal matrices $D_{m_\nu}=\mathrm{diag}(0,m_{\nu_2},m_{\nu_3})$ and $D_{m_N}=\mathrm{diag}(M_1,M_2)$. 
\subsection*{\ensuremath{\ell^\prime \to 3\ell} decay}
The branching ratio for $\ell^\prime \to 3\ell$ decay process is given by~\cite{Dinh:2013tvc}
\begin{eqnarray}
  \mathcal{B}R (\ell \to 3 \ell^\prime) = \frac{\alpha_{em}^2}{64 \pi^2 \sin^4{\theta_w}} |\mathcal{C}_{\ell3\ell^\prime}|^2 \times \mathcal{B}R(\ell\to\ell^\prime\nu_{\ell}\bar{\nu}_{\ell^\prime}),
\end{eqnarray}
where
\begin{eqnarray}
|\mathcal{C}_{\ell3\ell^\prime}|^2 &=& 2 \Big|\frac{1}{2}F_{Box}^{\ell\ell^\prime\ell^\prime\ell^\prime} + F_z^{\ell\ell^\prime} - 2\sin^2{\theta_w}\left(F_z^{\ell\ell^\prime} - F_\gamma^{\ell\ell^\prime}\right)    \Big|^2 + 4 \sin^4{\theta_w}\left|F_z^{\ell\ell^\prime} - F_\gamma^{\ell\ell^\prime}\right|^2   \nonumber \\
&+&  16 \sin^2{\theta_w} \text{Re} \left[\left(F_z^{\ell\ell^\prime}+\frac{1}{2}F_{Box}^{\ell\ell^\prime\ell^\prime\ell^\prime} \right)G_\gamma^{\ell\ell^{\prime\ast}}\right] - 48  \sin^4{\theta_w} \text{Re} \left[\left(F_z^{\ell\ell^\prime} - F_\gamma^{\ell\ell^\prime}\right)G_\gamma^{\ell\ell^{\prime\ast}}\right] \nonumber\\
&+& 32 \sin^4{\theta_w} |G_\gamma^{\ell\ell^\prime}|^2 \left(\log\frac{m_\ell^2}{m_{\ell^\prime}^2}-\frac{11}{4}\right),
\end{eqnarray}
where the expressions of $F_{Box}^{\ell\ell^\prime\ell^\prime\ell^\prime}, F_\gamma^{\ell\ell^\prime}, F_z^{\ell\ell^\prime}$ and $G_\gamma^{\ell\ell^\prime}$ are given in Ref.~\cite{Dinh:2013tvc}.

\subsection*{$\mu$ to $e$ conversion rate in a nucleus}
The $\mu-e$ conversion rate in a nucleus $N$ is given as
\begin{eqnarray}
\text{CR}(\mu N \to e N) 
&=& \frac{\alpha_{em}^5}{2 \pi^4 \sin^4{\theta_w}}\frac{Z_{\rm eff}^4}{Z}\left|F(-m_\mu^2)\right|^2G_F^2 m_\mu^5 \times \left|(RV)_{\mu k}^\ast(RV)_{e k}\right|^2 \left|\mathcal{C}_{\mu e}\right|^2,
\end{eqnarray}
where the loop factor is given by
\begin{eqnarray}
    \mathcal{C}_{\mu e} \approx Z\left(2 F_u^{(\mu e)}(x_k) + F_d^{(\mu e)}(x_k)\right) +\text{N}\left( F_u^{(\mu e)}(x_k) + 2 F_d^{(\mu e)}(x_k)\right)
\end{eqnarray}
with 
\begin{eqnarray}
   F_q^{(\mu e)}(x_k) &=& Q_q \sin^2{\theta_w}\left[F_\gamma(x_k)-F_z^{(\mu e)}(x_k)+G_\gamma(x_k)\right] \nonumber \\
   &+& \frac{1}{4}\left[2I_3 F_z^{(\mu e)}(x_k)+F_B^{(\mu e qq)}(x_k)\right],
\end{eqnarray}
 The functions $F_\gamma(x_k), F_z^{(\mu e)}(x_k), G_\gamma(x_k)$ and $F_B^{(\mu e qq)}(x_k)$ are given in Ref.~\cite{Dinh:2013tvc}.

\section{Reaction densities in Boltzmann evolution }\label{app:gamma_beexp}
We collect the expressions for the various reaction densities appearing in the Boltzmann equations (see Eqs.~\eqref{eq:boltz}). All quantities are defined in the plasma rest frame and follow the conventions of Ref.~\cite{Plumacher:1996kc}. For a generic two‑body scattering process $a+b \to i+j+\cdots$, the reaction density is given by
\begin{eqnarray}
\gamma(a+b\to i+j+\cdots) = \frac{T}{64\pi^4}\int_{s_{\text{min}}}^{\infty} ds\,
\hat{\sigma}(s)\,\sqrt{s}\,K_1\!\left(\frac{\sqrt{s}}{T}\right),
\end{eqnarray}
where $s$ is the centre‑of‑mass energy squared, $K_1$ the modified Bessel function, and $\hat{\sigma}(s)$ the reduced cross section. The latter is related to the usual cross section $\sigma(s)$ by
\begin{eqnarray}
\hat{\sigma}(s) = \frac{8}{s}\Bigl[(p_a\!\cdot\!p_b)^2 - M_a^2 M_b^2\Bigr]\sigma(s).
\end{eqnarray}
For a decay process $N_j \to H L $, the equilibrium reaction density simplifies to
\begin{eqnarray}
\label{eq:gamma_N}
\gamma_{N_j} = N_{N_j}^{\text{eq}}\,\frac{K_1(z)}{K_2(z)}\,\Gamma_{N_j},
\qquad z = \frac{M_1}{T},
\end{eqnarray}
with $N_{N_j}^{\text{eq}}$ the equilibrium number density of $N_j$,
\begin{eqnarray}
N_{N_j}^{\text{eq}} = \frac{g M_j^2 T}{2\pi^2}\,K_2(z),
\end{eqnarray}
and $\Gamma_{N_j}$ the total decay width of $N_j$. 
\subsection*{$\Delta L = 1$ scatterings\label{append_cHs}}
These processes involve the exchange of the Standard Model Higgs and contribute to the washout. The reduced cross sections for these processes are given below.

\textbf{$s$-channel process $N_j l \to \bar{t} q$ :}
\begin{eqnarray}
    \hat{\sigma}_{H,s}^j(s) = \frac{3\pi\alpha^2 m_t^2}{m_W^4\sin^4\theta_W}\,(m_D^\dagger m_D)_{jj}\,
\left(\frac{x_j - a_j}{x_j}\right)^2,
\end{eqnarray}
with $x_j = s/M_j^2$ and $a_j = 1$.

\textbf{$t$-channel process $N_j t \to \bar{l} q$ :}
\begin{eqnarray}
\hat{\sigma}_{H,t}^j(s) = \frac{3\pi\alpha^2 m_t^2}{m_W^4\sin^4\theta_W}\,(m_D^\dagger m_D)_{jj}\,
\left[\frac{x_j - a_j}{x_j} + \frac{a_j}{x_j}\ln\!\left(\frac{x_j - a_j + y'}{y'}\right)\right],
\end{eqnarray}
where $y' = m_h^2/M_j^2$ and $m_h$ is the Higgs mass.
\subsection*{$\Delta L = 2$ scatterings\label{append_cNs}}
Heavy neutrinos mediate these processes and contribute to the washout; their reduced cross-sections are:

\textbf{$s$-channel process $l l \to H H$:}
\begin{eqnarray}
\hat{\sigma}_{N,s}(s)& =& \frac{\alpha^2}{\sin^4\theta_W} \frac{2\pi}{m_W^4} \frac{1}{x} 
\Biggl\{ \sum_{j=1}^2 a_j (m_D^\dagger m_D)_{jj}^2 
\Biggl[ \frac{x}{a_j} + \frac{2x}{D_j(x)} + \frac{x^2}{2D_j^2(x)} \nonumber\\
& - &\left(1 + 2\frac{x + a_j}{D_j(x)}\right) \ln\left(\frac{x + a_j}{a_j}\right) \Biggr]\nonumber \\
&+& 2\sqrt{a_1 a_2}\, \text{Re}\!\left[(m_D^\dagger m_D)_{12}^2\right]
\Biggl[ \frac{x}{D_1(x)} + \frac{x}{D_2(x)} + \frac{x^2}{2D_1(x)D_2(x)} \nonumber\\
& -& \frac{(x + a_1)(x + a_1 - 2a_2)}{D_2(x)(a_1 - a_2)} \ln\left(\frac{x + a_1}{a_1}\right) \nonumber\\
& -& \frac{(x + a_2)(x + a_2 - 2a_1)}{D_1(x)(a_2 - a_1)} \ln\left(\frac{x + a_2}{a_2}\right) \Biggr] \Biggr\},
\end{eqnarray}
where $x = s/M_1^2$, $a_j = 1$, and
\begin{eqnarray}
   \frac{1}{D_j(x)} := \frac{x - a_j}{(x - a_j)^2 + a_j c_j}, \qquad
c_j := \left(\frac{\Gamma_{N_j}}{M_j}\right)^2. 
\end{eqnarray}
The prime indicates subtraction of real intermediate states.

\textbf{$t$-channel process $l H \to \bar l H^\dagger$:}
\begin{eqnarray}
\hat{\sigma}_{N,t}(s) &&= \frac{2\pi\alpha^2}{m_W^4\sin^4\theta_W} 
\Biggl\{ \sum_{j=1}^2 a_j (m_D^\dagger m_D)_{jj}^2 
\Biggl[ \frac{1}{2a_j}\frac{x_j}{x_j + a_j} + \frac{1}{x_j + 2a_j}\ln\left(\frac{x_j + a_j}{a_j}\right) \Biggr] \nonumber\\
&&+ \text{Re}\!\left[(m_D^\dagger m_D)_{12}^2\right] 
\frac{\sqrt{a_1 a_2}}{(a_1 - a_2)(x_1 + a_1 + a_2)} \nonumber\\
&&\times \Biggl[ (x_1 + 2a_1)\ln\left(\frac{x_1 + a_2}{a_2}\right) 
- (x_1 + 2a_2)\ln\left(\frac{x_1 + a_1}{a_1}\right) \Biggr] \Biggr\},
\end{eqnarray}
with $x_j = s/M_j^2$. 

\label{Bibliography}
\bibliographystyle{JHEP}
\bibliography{Refs}

\end{document}